\documentclass[preprint]{aastex631}
\shorttitle{large sCMOS sensor performance}
\shortauthors{Wu et al.}
\graphicspath{{./}{figure/}}
\begin{document}

\title{X-ray performance of a customized large-format scientific CMOS detector}

\author{Qinyu Wu}
\affiliation{National Astronomical Observatories, Chinese Academy of Sciences \\
20A Datun Road, Chaoyang District \\
Beijing 100101, China}
\affiliation{School of Astronomy and Space Science, University of Chinese Academy of Sciences \\
19A Yuquan Road, Shĳingshan District \\
Beijing 100049, China}

\author{Zhenqing Jia}
\affiliation{National Astronomical Observatories, Chinese Academy of Sciences \\
20A Datun Road, Chaoyang District \\
Beijing 100101, China}

\author{Wenxin Wang}
\affiliation{National Astronomical Observatories, Chinese Academy of Sciences \\
20A Datun Road, Chaoyang District \\
Beijing 100101, China}

\author{Zhixing Ling}
\affiliation{National Astronomical Observatories, Chinese Academy of Sciences \\
20A Datun Road, Chaoyang District \\
Beijing 100101, China}
\affiliation{School of Astronomy and Space Science, University of Chinese Academy of Sciences \\
19A Yuquan Road, Shĳingshan District \\
Beijing 100049, China}
\correspondingauthor{Zhixing Ling}
\email{lingzhixing@nao.cas.cn}

\author{Chen Zhang}
\affiliation{National Astronomical Observatories, Chinese Academy of Sciences \\
20A Datun Road, Chaoyang District \\
Beijing 100101, China}
\affiliation{School of Astronomy and Space Science, University of Chinese Academy of Sciences \\
19A Yuquan Road, Shĳingshan District \\
Beijing 100049, China}

\author{Shuangnan Zhang}
\affiliation{National Astronomical Observatories, Chinese Academy of Sciences \\
20A Datun Road, Chaoyang District \\
Beijing 100101, China}
\affiliation{School of Astronomy and Space Science, University of Chinese Academy of Sciences \\
19A Yuquan Road, Shĳingshan District \\
Beijing 100049, China}
\affiliation{Institute of High Energy Physics, Chinese Academy of Sciences \\
19B Yuquan Road, Shĳingshan District \\
Beijing 100049, China}

\author{Weimin Yuan}
\affiliation{National Astronomical Observatories, Chinese Academy of Sciences \\
20A Datun Road, Chaoyang District \\
Beijing 100101, China}
\affiliation{School of Astronomy and Space Science, University of Chinese Academy of Sciences \\
19A Yuquan Road, Shĳingshan District \\
Beijing 100049, China}



\begin{abstract}

In recent years, the performance of Scientific Complementary Metal Oxide Semiconductor (sCMOS) sensors has been improved significantly. Compared with CCD sensors, sCMOS sensors have various advantages, making them potentially better devices for optical and X-ray detection, especially in time-domain astronomy. After a series of tests of sCMOS sensors, we proposed a new dedicated high-speed, large-format X-ray detector in 2016 cooperating with Gpixel Inc. This new sCMOS sensor has a physical size of 6 cm $\times$ 6 cm, with an array of $4096\times4096$ pixels and a pixel size of $\rm{15\ \mu m}$. The frame rate is 20.1 fps under current condition and can be boosted to a maximum value around 100 fps. The epitaxial thickness is increased to $\rm{10\ \mu m}$ compared to the previous sCMOS product. We show the results of its first taped-out product in this work. The dark current of this sCMOS is lower than 10 $\rm{e^-}$/pixel/s at $\rm{20 ^{\circ}\!C}$, and lower than 0.02 $\rm{e^-}$/pixel/s at $\rm{-30 ^{\circ}\!C}$. The Fixed Pattern Noise (FPN) and the readout noise are lower than 5 $\rm{e^-}$ in high-gain situation and show a small increase at low temperature. The energy resolution reaches 180.1 eV (3.1\%) at 5.90 keV for single-pixel events and 212.3 eV (3.6\%) for all split events. The continuous X-ray spectrum measurement shows that this sensor is able to response to X-ray photons from 500 eV to 37 keV. The excellent performance, as demonstrated from these test results, makes sCMOS sensor an ideal detector for X-ray imaging and spectroscopic application.

\end{abstract}

\keywords{X-ray detectors (1815), Astronomical instrumentation (799), Astronomical detectors (84)}


\section{Introduction}
\label{sec:introduction}
In the past several decades, the traditional Charge-coupled Device (CCD) has dominated in optical and X-ray astronomical applications. Many successful X-ray missions have chosen CCD sensors as their focal plane detectors, including ASCA \citep{burke1994ccd}, Chandra \citep{garmire2003advanced}, XMM-Newton \citep{struder2001european} and eROSITA \citep{meidinger2010development}. High-speed, large-format sensors are required for the next generation of large-area X-ray missions, e.g., the Lynx mission \citep{gaskin2019lynx}. In recent years, the performance of Scientific Complementary Metal Oxide Semiconductor (sCMOS) detectors has been improved considerably. Compared with CCD detectors, sCMOS detectors have a number of advantages: high readout frame rate, no charge transfer, high circuit integration, high radiation tolerance, low readout noise and high working temperature, making them potentially promising devices for optical and X-ray applications, especially in time-domain astronomy and wide-field sky surveys. Several X-ray missions and mission concepts employ sCMOS sensors, such as FOXSI3 \citep{ishikawa2018high} for solar X-ray observation, Einstein Probe (EP) and THESEUS \citep{heymes2020development} for the detection of X-ray transient sources. In the 2000s, hybrid CMOS sensors were first studied for X-ray detection \citep{falcone2007hybrid}. In the following years, sCMOS sensors have sprung up, including monolithic sCMOS sensors and hybrid sCMOS sensors. Teledyne has released a number of sCMOS products, including the CIS115 sensor designed for the Jupiter Icy Moon Explorer (JUICE) \citep{soman2014design}, and the newly released COSMOS astronomy camera with an 8 cm large-format sensor\footnote{\url{https://www.princetoninstruments.com/products/cosmos-family/cosmos}}. Gpixel Inc. has also released a number of high performance sCMOS sensors since the release of GSENSE\-400\-BSI in 2015 \citep{ma2015a}. For soft X-ray detection, a sCMOS sensor was produced with a 5-nm-thick entrance layer, reaching a quantum efficiency (QE) larger than 90\% in the photon-energy range of 80-1000 eV \citep{harada2020high}. 

We have been studying the X-ray performance of sCMOS sensors and their applications in X-ray astronomy based on GSENSE\-400\-BSI since 2015. We have proved that sCMOS sensors are feasible for X-ray astronomical observations \citep{wang2019characterization, ling2021correlogram}. Cooperating with Gpixel Inc., we proposed an X-ray sCMOS detector with enlarged format and pixel size, and a thickened epitaxial layer in 2016. The final design was completed in 2018. And the first samples of the new sensor, GSENSE\-1516\-BSI, were delivered in 2019. A series of tests of the basic properties of the sensor have been carried out, showing an excellent performance matching the expectation.

The basic properties of the GSENSE\-1516\-BSI sCMOS sensor are described in section \ref{basic_property}; the measurements of X-ray performance are presented in section \ref{xray_performance}; and conclusions are summarized in section \ref{conclusions}.

%--------------------------------------------------------------------
\section{Basic properties of GSENSE\-1516\-BSI}

\begin{figure}[htbp]
\centering
\resizebox{0.8\hsize}{!}{\includegraphics{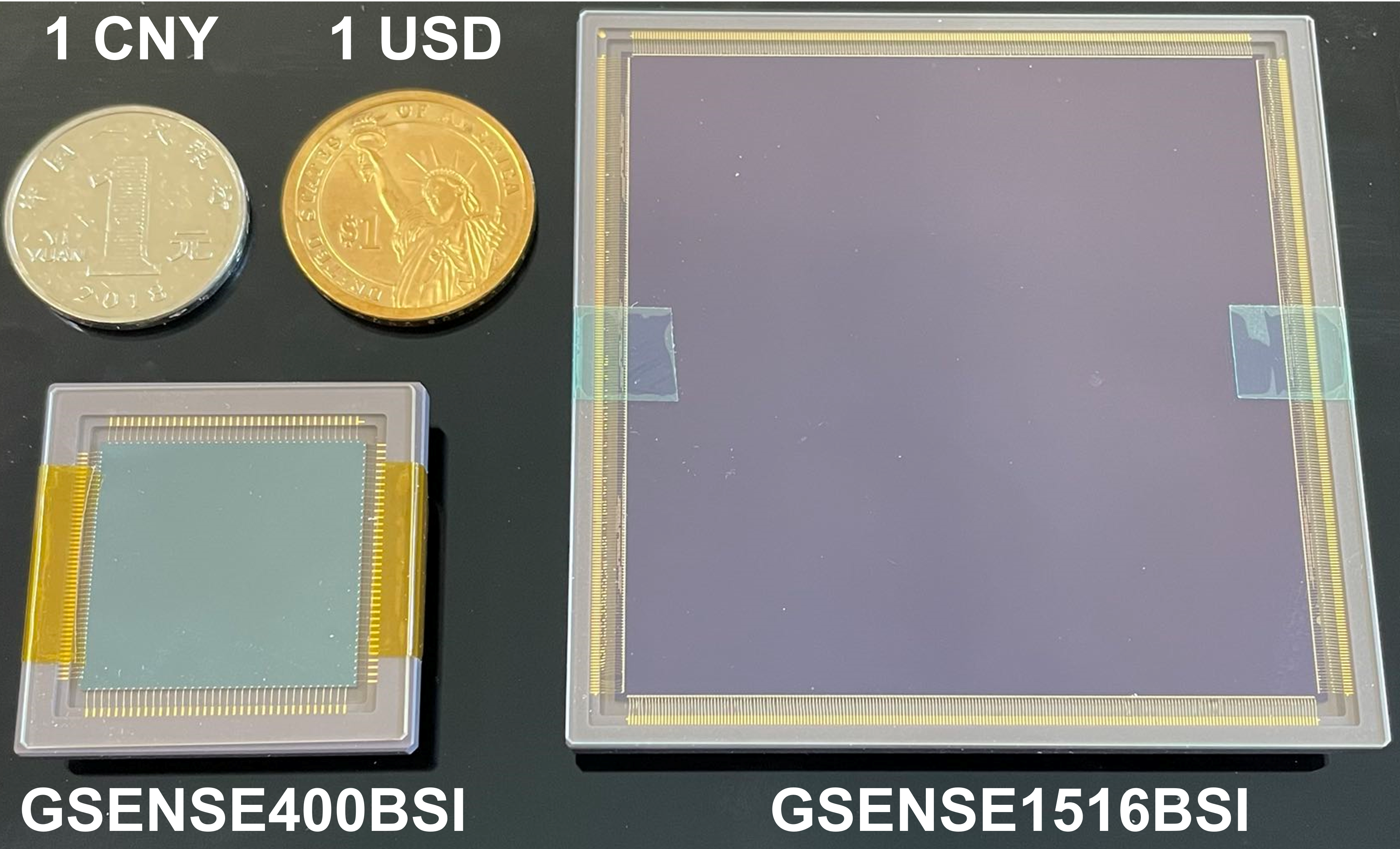}}
\caption{A GSENSE\-1516\-BSI sCMOS sensor (right) and a GSENSE\-400\-BSI sCMOS sensor (left). The much larger image area of GSENSE\-1516\-BSI makes it an excellent sensor for X-ray detection.}
\label{CMOS_photo}
\end{figure}

%-------------------------------------- Two column figure (place early!)
%   \begin{figure*}
%   \centering
%   %%%\includegraphics{empty.eps}
%   %%%\includegraphics{empty.eps}
%   %%%\includegraphics{empty.eps}
%   \caption{Adiabatic exponent $\Gamma_1$.
%               $\Gamma_1$ is plotted as a function of
%               $\lg$ internal energy $\mathrm{[erg\,g^{-1}]}$ and $\lg$
%               density $\mathrm{[g\,cm^{-3}]}$.}
%               \label{FigGam}%
%     \end{figure*}
%
  \label{basic_property}
As a back-illuminated scientific CMOS sensor, GSENSE\-1516\-BSI (Fig. \ref{CMOS_photo}) has an effective pixel array of $4096\times 4096$ with a pixel size of $15\ \rm{\mu m}\times 15\ \rm{\mu m}$, reaching a total image area of 6 cm $\times$ 6 cm. The frame rate of this sensor is 20.1 fps under current condition and can be boosted to a maximum value around 100 fps. Its basic properties are summarized in Table \ref{table_pro}. 

\begin{table}[h!]
\centering
\resizebox{\hsize}{!}{
\begin{tabular}{| l | l |}
\hline
Image area & $6\ \rm{cm}\times 6\ \rm{cm}$ \\ \hline
Pixel size & $15\ \rm{\mu m}\times 15\ \rm{\mu m}$ \\ \hline
Epitaxial thickness & $10\ \rm{\mu m}$ \\ \hline
%Total pixels & $4154\times 4112$ \\ \hline
Number of pixels & $4096\times 4096$ \\ \hline
Frame rate & 20.1 fps (current) / $\sim$100 fps (maximum) \\ \hline
ADC digit & 12 bit \\ \hline
Shutter format & Rolling shutter \\ \hline
Physical full well capacity (FWC) & 120K$\rm{e^-}$ \\ \hline
%Dynamic range & 69.11 dB \\ \hline
Photo-response non-uniformity (PRNU) & $<2\%$ \\ \hline
Fixed pattern noise (FPN) & $\rm{< 5.0\ e^-}$ at high gain\\ \hline
Readout noise & $\rm{< 5.0\ e^-}$ at high gain\\ \hline
%Fixed pattern noise (FPN) & $\rm{\sim 3.9e^-\  @\  20^{\circ}\!C}$ \\ \hline
%Readout noise & $\rm{\sim 3.3e^-\  @\  20^{\circ}\!C}$\\ \hline
Dark current & $\rm{< 0.02\ e^-/pixel/s\  @\ -30^{\circ}\!C}$ \\ 
& $\rm{< 10\ e^-/pixel/s\  @\  20^{\circ}\!C}$ \\ \hline
Supply voltage & 5 V (analog) / 1.8 V (digital) \\ \hline
Power consumption & $<1.6 \rm{W}$ \\ \hline
\hline
\end{tabular}}
\caption{Basic properties of GSENSE\-1516\-BSI.}
\label{table_pro}
\end{table}

\begin{figure}[htbp]
\centering
\resizebox{0.8\hsize}{!}{\includegraphics{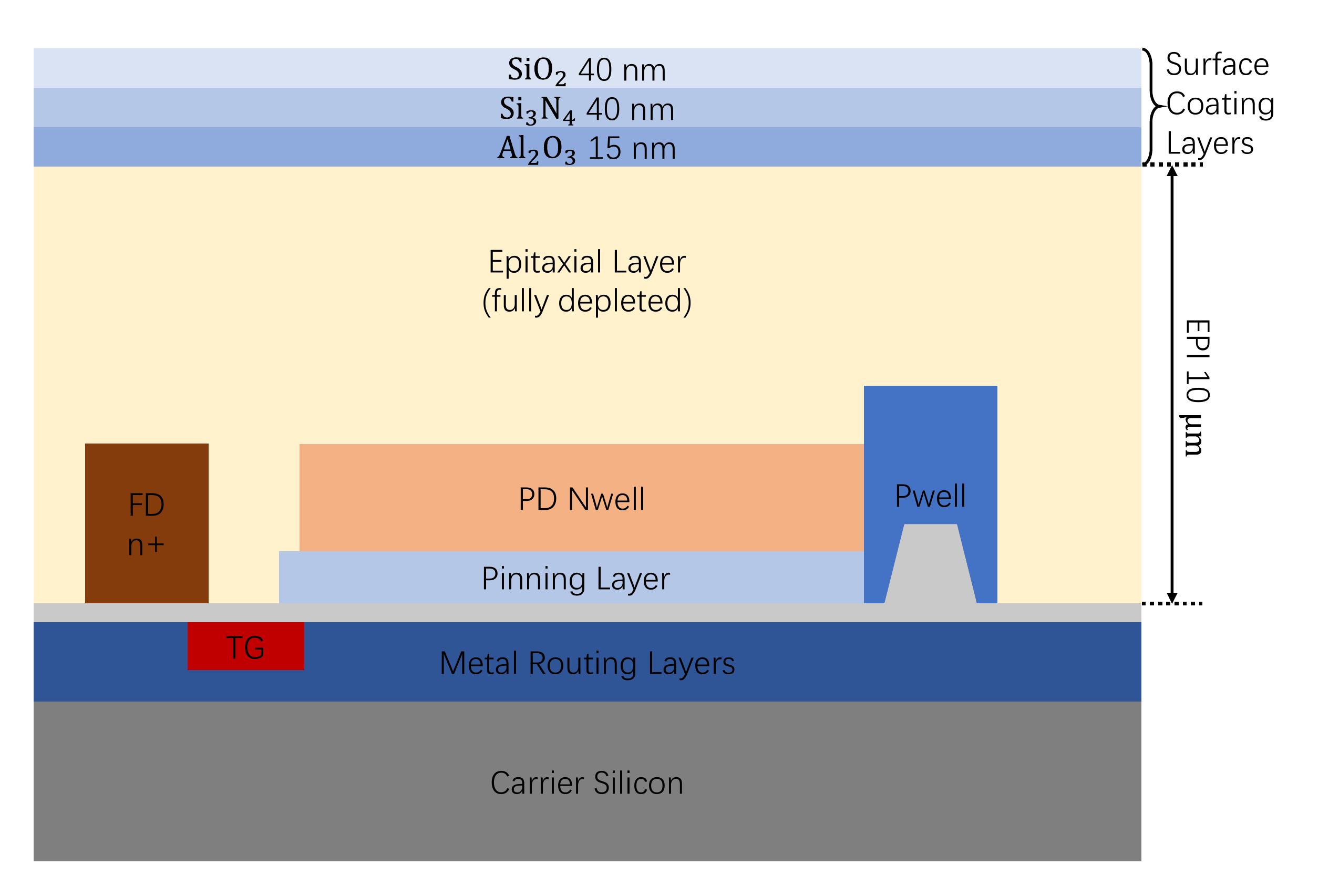}}
\caption{Illustration of the pixel structure of GSENSE\-1516\-BSI. (credit: Gpixel Inc.)}
\label{CMOS_structure}
\end{figure}

Both GSENSE\-1516\-BSI and its predecessor GSENSE\-400\-BSI are aiming at high sensitivity applications. An illustration of the pixel structure of the new sensor is given in Fig. \ref{CMOS_structure}. Compared to GSENSE\-400\-BSI, the new sensor has several design improvements, listed below: 
\begin{itemize}
\item GSENSE\-1516\-BSI uses a thicker high resistivity Epitaxial layer to increase the depletion depth. It helps to reduce the inter-pixel crosstalk.
\item Due to the larger pixel pitch, the PD area of the new sensor is much larger, which is beneficial for both crosstalk performance and full well capacity (FWC).
\item Considering the larger PD area, we adopted a smaller $\rm{V_{pin}}$ to decrease the lag risk. This can improve the photo-response linearity under weak illumination.
\item GSENSE\-1516\-BSI uses a 5 V rather than a 3.3 V operation voltage. This higher voltage can increase the floating diffusion (FD) voltage swing and the dynamic range.
\end{itemize}

\begin{figure}[htbp]
\centering
\resizebox{0.6\hsize}{!}{\includegraphics{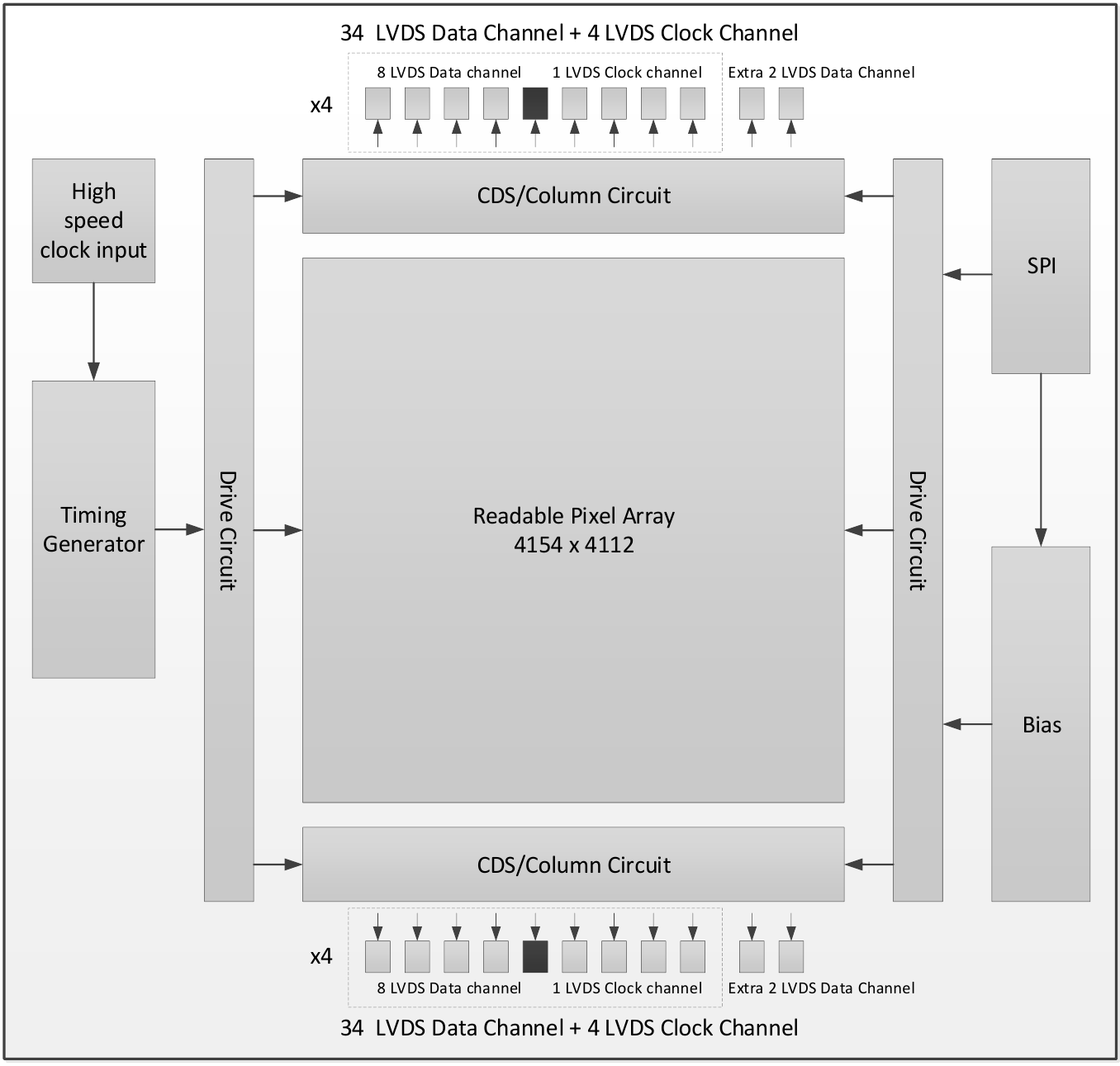}}
\caption{Illustration of the structure of GSENSE\-1516\-BSI. (From the official datasheet by Gpixel Inc.)}
\label{sensor_structure}
\end{figure}

The structure of the GSENSE\-1516\-BSI sensor is shown in Fig. \ref{sensor_structure}. Each column in the sensor has its own CDS circuit, amplifier and ADC. Currently, 34 pairs of LVDS data channels and a clock of 130 MHz are used to achieve a frame rate of 20.1 fps. Indeed, the sensor has 68 pairs of LVDS channels and has the capability of running under a 300 MHz clock, which makes it possible to reach a maximum frame rate around 100 fps. 

\begin{figure}[htbp]
\centering
\resizebox{\hsize}{!}{
\includegraphics[width=0.45\textwidth]{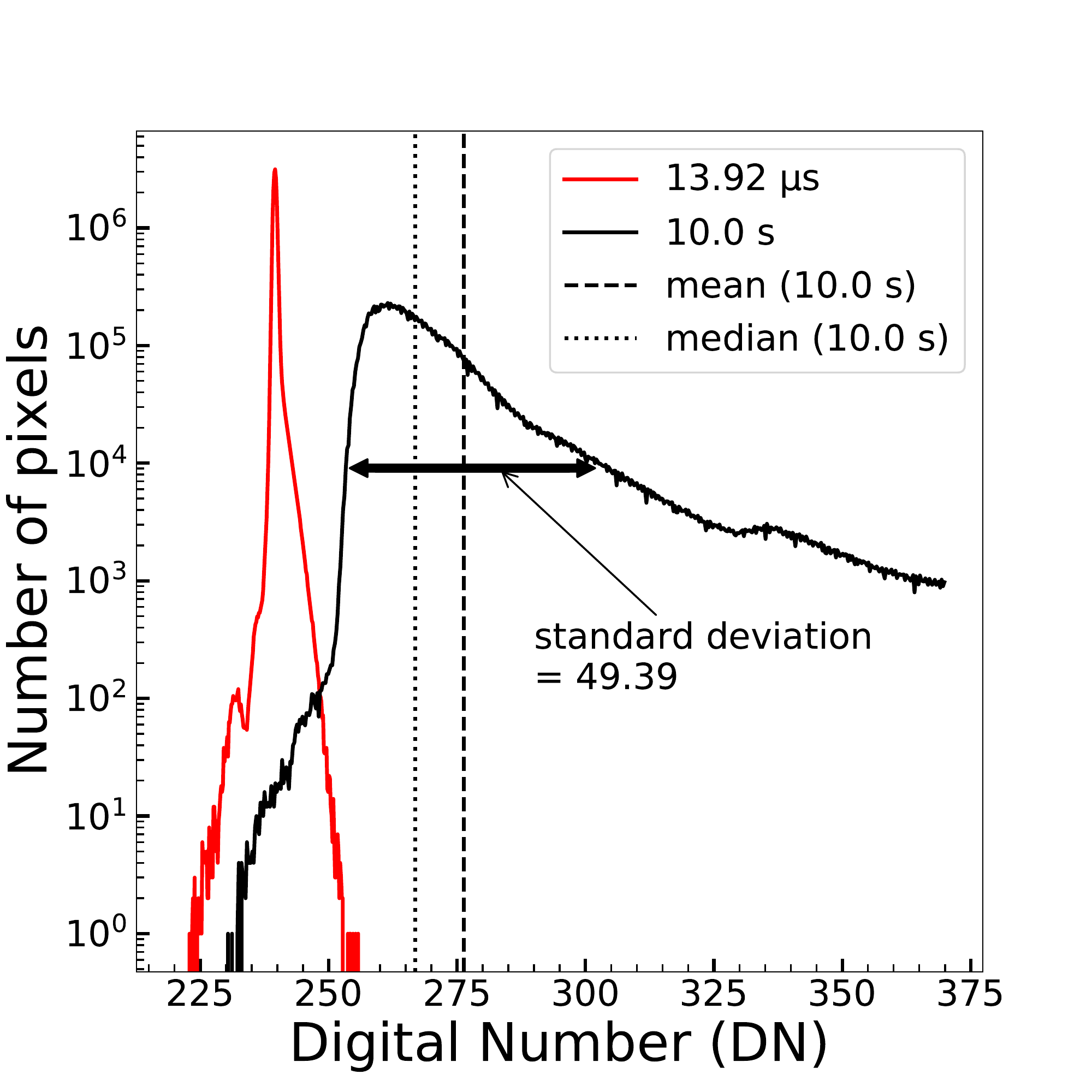}
%\hspace{1in}
\includegraphics[width=0.45\textwidth]{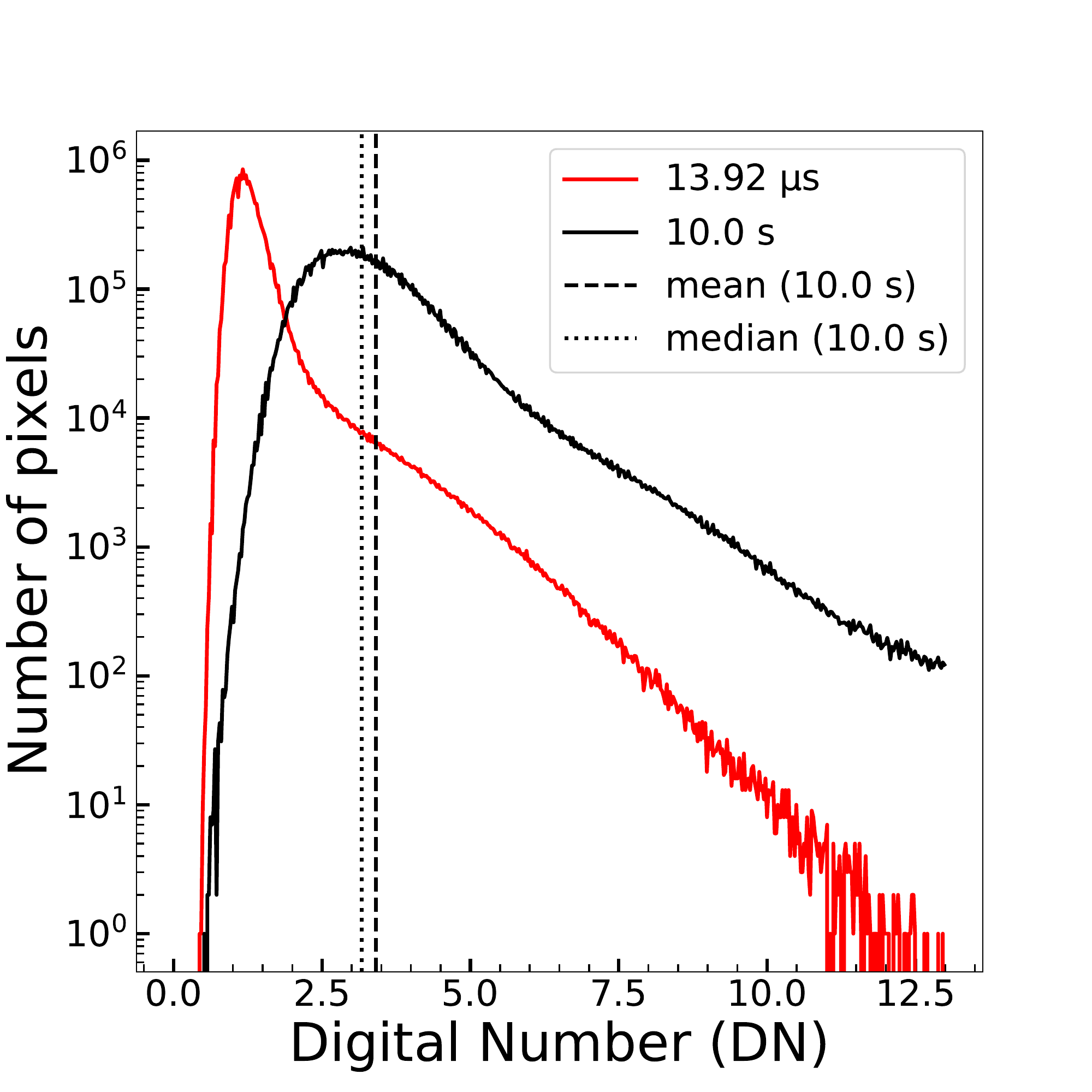}}
\caption{The distribution of the bias (left) and the noise (right) among pixels of the GSENSE\-1516\-BSI at $\rm{20 ^{\circ}\!C}$. The left panel shows the histogram of the bias map at $\rm{\sim 13.92\ \mu s}$ exposure (red), and the histogram at 10 s exposure (black). The right panel shows the distribution of their standard deviations. The dashed and dotted vertical lines indicate the mean and median of the distribution for the 10 second exposure.}
\label{fig_compare}
\end{figure}

\begin{figure}[htbp]
\centering
\resizebox{\hsize}{!}{
\includegraphics[width=0.53\textwidth]{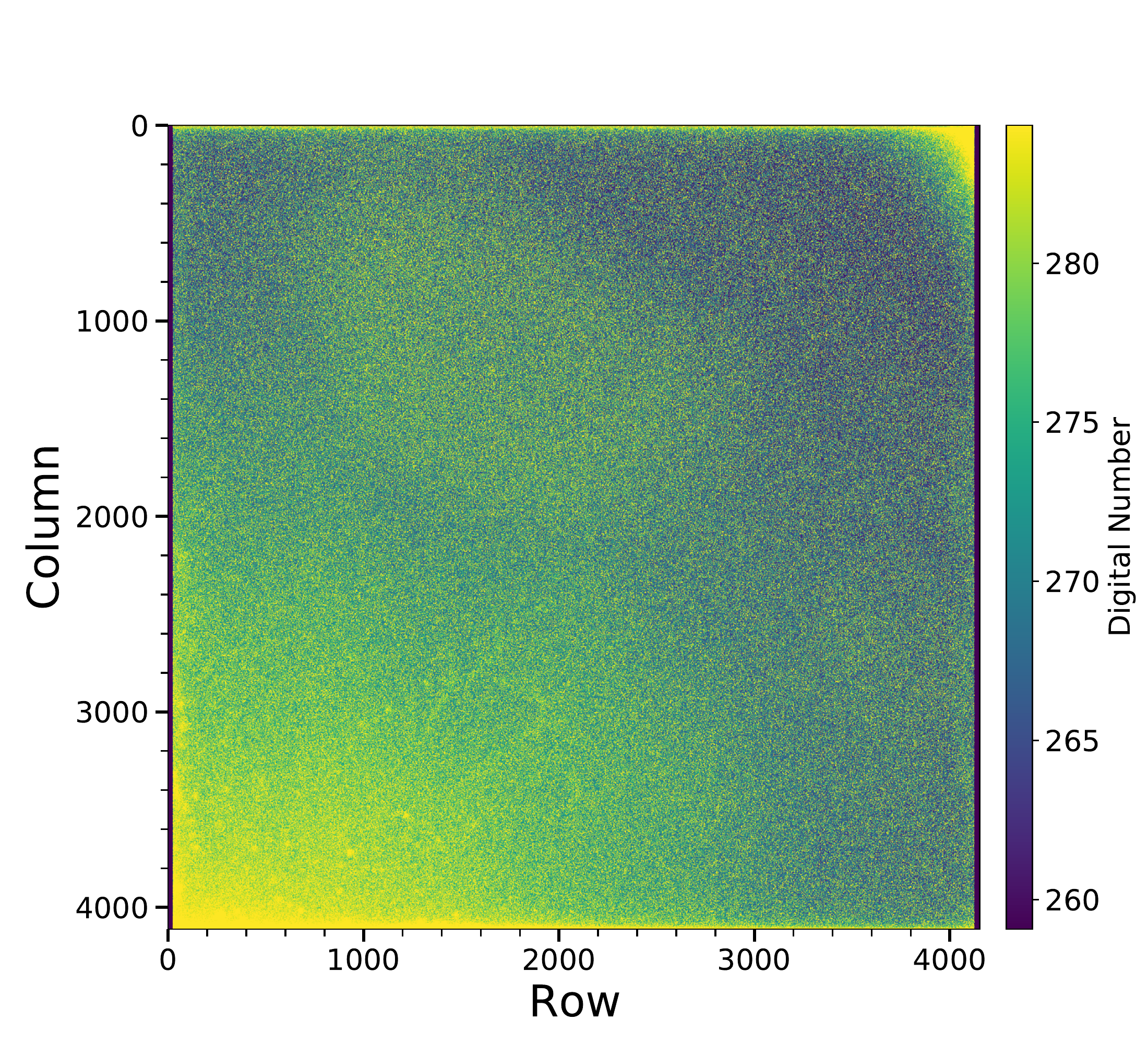}
%\hspace{1in}
\includegraphics[width=0.53\textwidth]{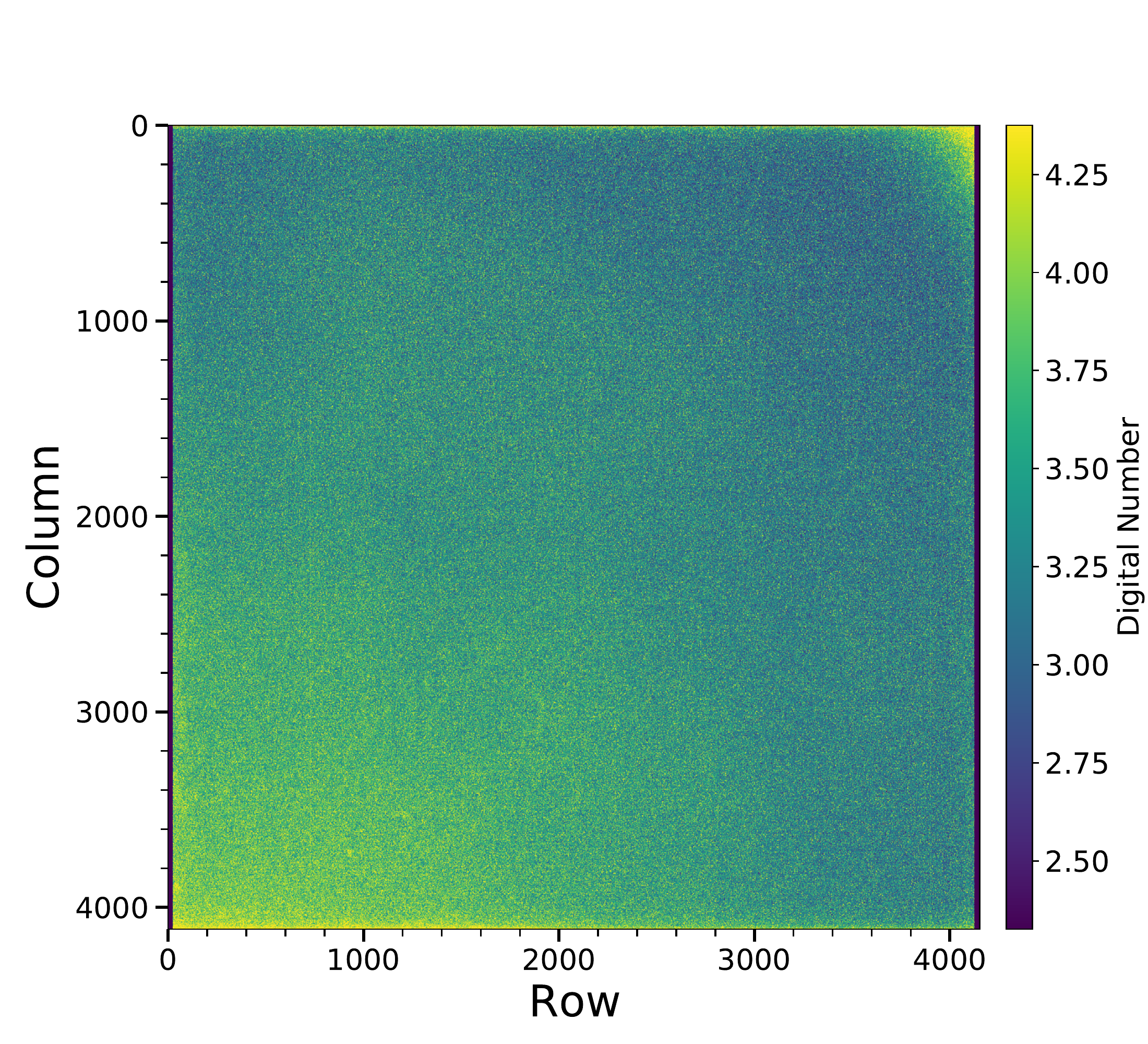}}
\caption{The bias map (left panel) and the noise map (right panel) with a 10 second exposure at $\rm{20 ^{\circ}\!C}$. These are the 2-dimensional maps corresponding to the two histograms (black) in Fig. \ref{fig_compare}. The dark current non-uniformity (partly represented by the bias map) and the variation of readout noise among pixels (partly represented by the noise map) can be seen clearly.}
\label{fig_map}
\end{figure}

The fixed pattern noise (FPN), the readout noise and the dark current of the sensor were measured in a dark vacuum environment at a given temperature controlled by a peltier cooler. A mean image is calculated from thirty dark frames with the shortest integration time ($\rm{\sim 13.92\ \mu s}$).  The red line in the left panel of Fig. \ref{fig_compare} shows the distribution of the mean values of all pixels. The standard deviation of this distribution is 1.45 DN, corresponding to the FPN. The pixel-by-pixel standard deviations over these 30 dark frames are plotted in a histogram, shown as the red line in the right panel of Fig. \ref{fig_compare}. Given the highly skewed distribution with a significant tail, we use the median rather than the mean to represent the readout noise, around 1.24 DN. This tail is due to the non-uniformity among pixels, which also has been found in GSENSE\-400\-BSI \citep{ling2021correlogram}. As for the dark current measurement, several dark images were recorded with an exposure time, and this process was repeated for a range of integration times from $\rm{\sim 13.92\ \mu s}$ to $100\ \rm{s}$ at a fixed temperature. The distributions of the mean values and their standard deviations for each of the pixels with a 10 s integration time are overplotted in the left and right panel in Fig. \ref{fig_compare}, respectively (black). The dark current of each pixel was obtained by applying a linear fit to the dark charges over a range of integration times. Then the median value among all the pixels is used to represent the dark current level of the sensor. The dark current and the readout noise are found to vary from pixel to pixel, as illustrated in Fig. \ref{fig_map}. The chip tested in this work shows an obvious large-region non-uniformity: the corner region are brighter than the rest of the chip. This non-uniformity pattern can vary chip by chip, which may come from the imperfection in the manufacturing process. However, this non-uniformity will diminish rapidly as the temperature decreases.

\begin{figure}[htbp]
\centering
\resizebox{\hsize}{!}{
\includegraphics{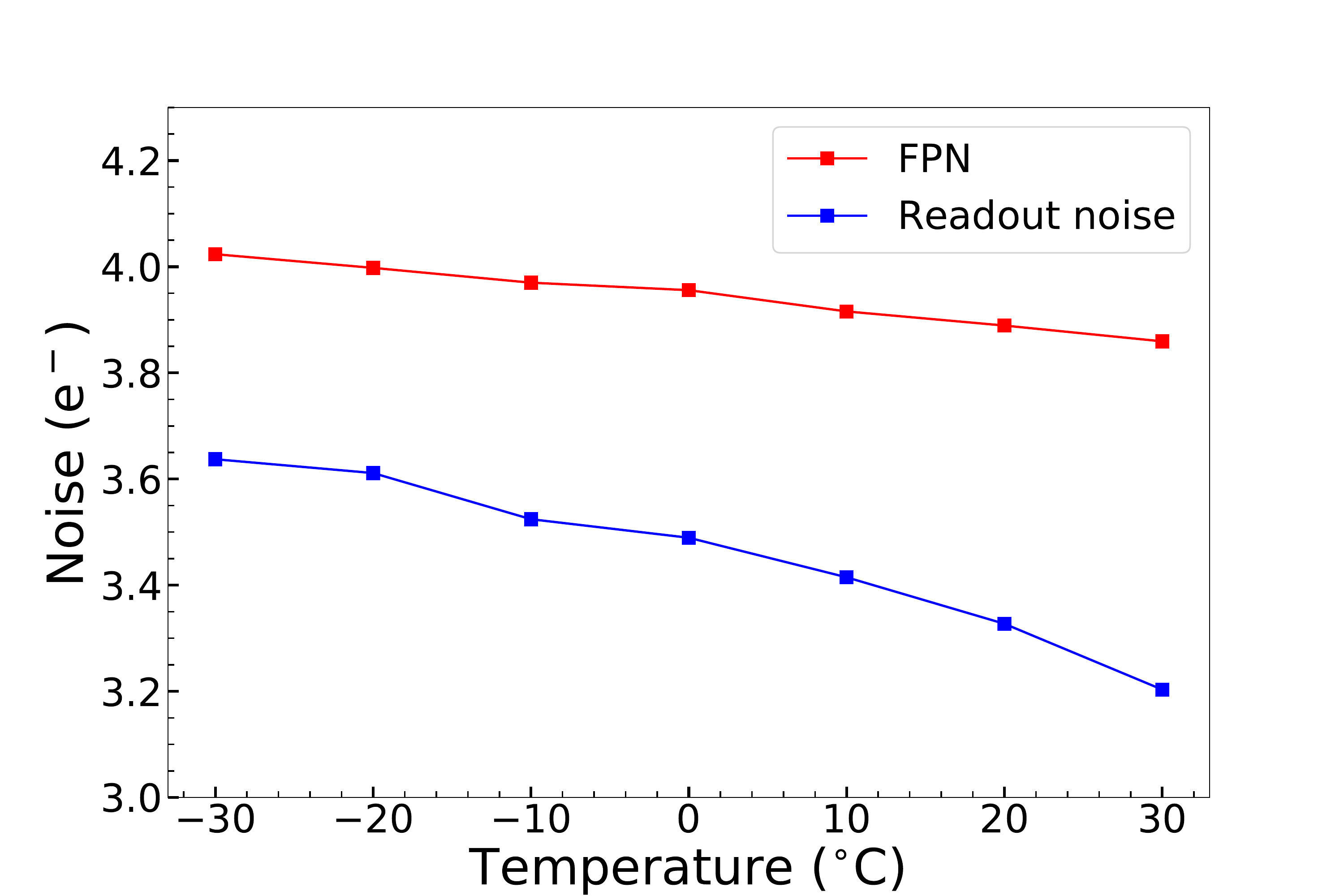}}
\caption{The dependence of the FPN (red) and the readout noise (blue) on temperature. At $\rm{-30 ^{\circ}\!C}$, the readout noise is approximately $3.64\ e^-$ and the FPN around $4.02\ e^-$. Both of them increase slightly at lower temperatures.}
\label{fig_noise_temp}
\end{figure}

\begin{figure}[htbp]
\centering
\resizebox{\hsize}{!}{
\includegraphics{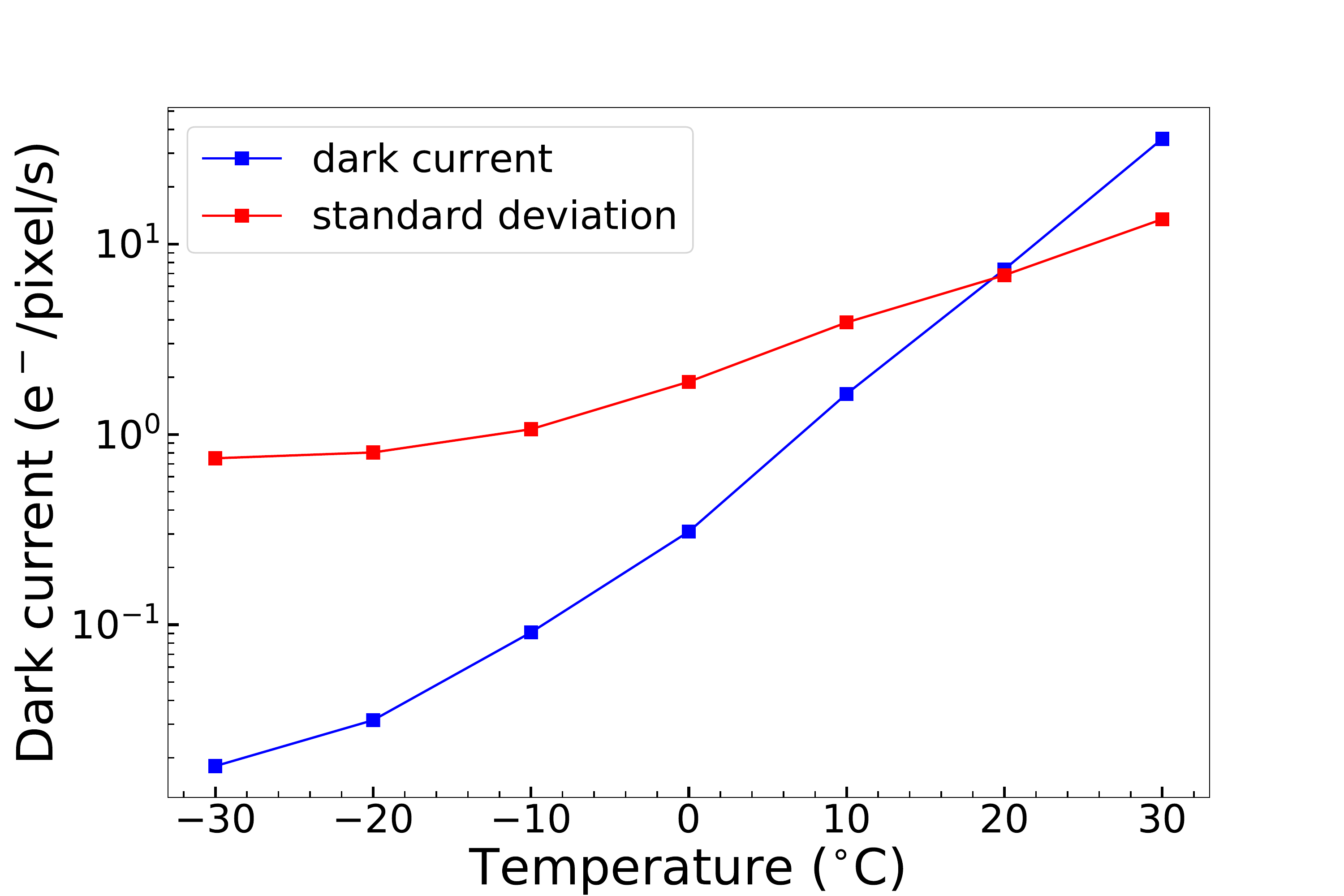}}
\caption{The dependence of the dark current (blue) and its standard deviation among pixels(red) on temperature. The dark current is represented by the median value among pixels. The dark current is as low as 0.018 $\rm{e^-}$/pixel/s at $\rm{-30 ^{\circ}\!C}$ and increases dramatically when temperature increases. The standard deviation of the dark current also increases with temperatures, but with a less steep slope.}
\label{dc_vs_temp}
\end{figure}

We measured these parameters at several different temperatures, ranging from $\rm{-30 ^{\circ}\!C}$ to $\rm{30 ^{\circ}\!C}$. At $\rm{20 ^{\circ}\!C}$, the readout noise is approximately $3.3\ e^-$ and the FPN is around $3.9\ e^-$. Fig. \ref{fig_noise_temp} shows that both of them increase slightly at lower temperatures. This temperature dependence is consistent with the result of other sCMOS sensors from Gpixel Inc.\citep{wang2019characterization}. Typically, transistors run faster at lower temperature. The speed increase at low temperature, although not utilized in these sensors, can result in a slightly higher noise level. The dark current is caused by random thermal excitation within the depletion region. As shown in Fig. \ref{dc_vs_temp}, the dark current is 7.4 $\rm{e^-}$/pixel/s at $\rm{20 ^{\circ}\!C}$ and decreases dramatically when temperature decreases, reaching 0.018 $\rm{e^-}$/pixel/s at $\rm{-30 ^{\circ}\!C}$. To quantify the non-uniformity among pixels, we introduced the standard deviation of the distribution of the pixel-by-pixel dark currents. As shown in Fig. \ref{dc_vs_temp}, a higher temperature results in not only a larger dark current but also a larger variation between pixels. This variation should be treated carefully in long-time exposures. The conversion gain adopted here is derived in Section \ref{sec_gain}.

This sCMOS sensor can also be used for optical detection. The quantum efficiency (QE), photo-response non-uniformity (PRNU) and other parameters can be found in the official datasheet (contact Gpixel Inc.\footnote{\url{https://www.gpixel.com/}} for the datasheet), and are not shown here.

%
%                                                One column figure
%----------------------------------------------------------------- 
%   \begin{figure}
%   \centering
%   %%%\includegraphics[width=3cm]{empty.eps}
%       \caption{Vibrational stability equation of state
%               $S_{\mathrm{vib}}(\lg e, \lg \rho)$.
%               $>0$ means vibrational stability.
%               }
%          \label{FigVibStab}
%   \end{figure}
%-----------------------------------------------------------------

\section{X-ray performance of GSENSE\-1516\-BSI}
\label{xray_performance}
Several X-ray sources were used to test the performance of GSENSE\-1516\-BSI. Firstly, we used an $^{55}\!\rm{Fe}$ X-ray source to test the energy response. Then the response to the continuous energy spectrum was studied with a Ag targert X-ray tube source from AMPTEK. Finally, an X-ray-illuminated Mg target was applied for the test of low energy response.

\subsection{Data extraction method}
\begin{figure}[htbp]
\centering
\resizebox{\hsize}{!}{
\includegraphics{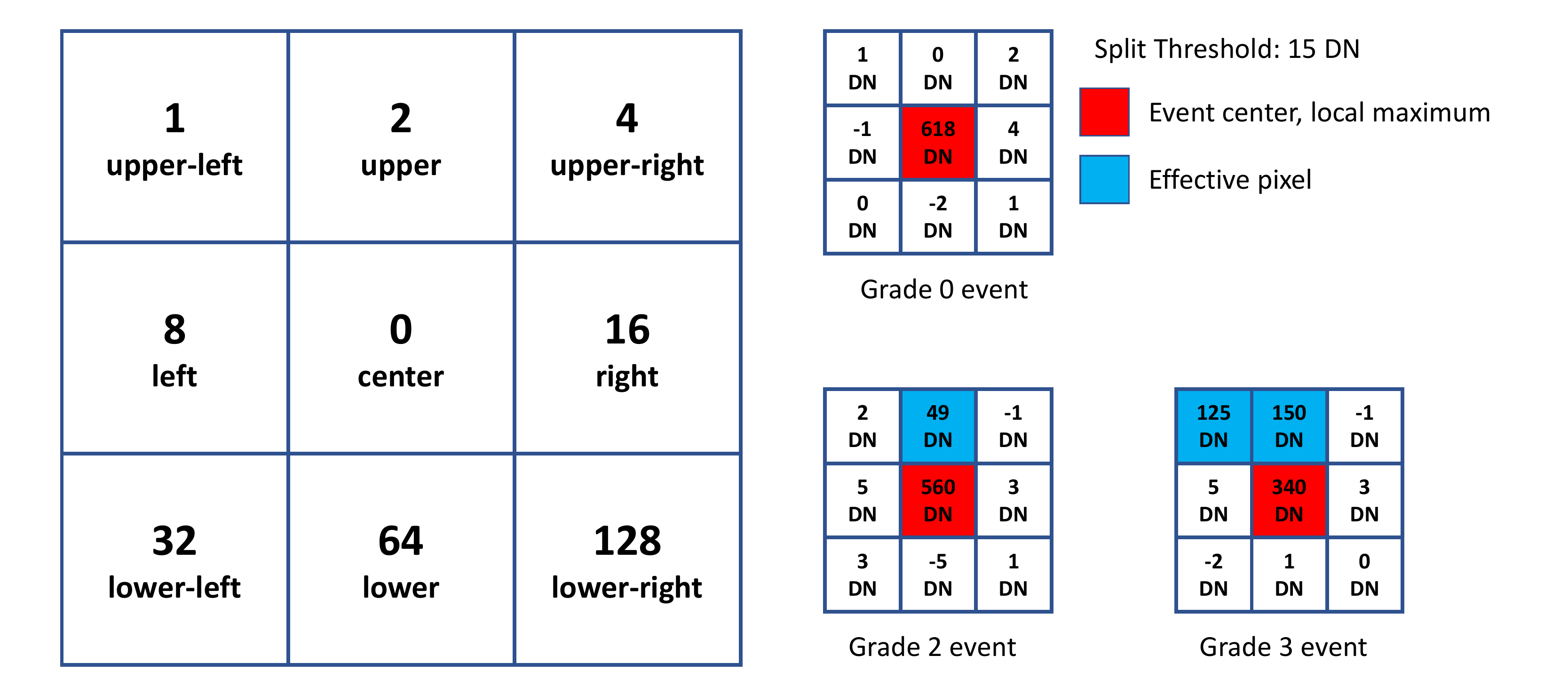}}
\caption{Grade definition of an X-ray event (Redrawing from Figure 4 of \citet{ling2021correlogram}). The left panel shows the definition of the event grade. The right panel gives some examples of Grade 0, Grade 2 and Grade 3 events. The $T_{split}$ is set to 15 DN.}
\label{grades_def}
\end{figure}

The data processing method is similar to that in \citet{ling2021correlogram} and is only summarized briefly here. A bias map is obtained from the first fifty raw images. Then the bias image is subtracted from all the raw images. After that, these images are searched over for X-ray events. Each pixel's digital number is compared to the event threshold, which is set to $T_{event}=100$ DN ($\sim$ 1 keV) for the $^{55}\!\rm{Fe}$ source. If an over-threshold pixel is the local maximum among its adjacent $3\times3$ pixels, the digital numbers in this $3\times3$ region are recorded as an event. A region of $3\times3$ pixels is sufficient to collect all the electrons in the cloud induced by a single soft X-ray photon for the GSENSE\-1516\-BSI sensor. One event can have a different distribution of the data pattern from another. A grade is assigned to each X-ray event extracted to characterize its split pattern, by using the ACIS (Advanced CCD Imaging Spectrometer) grade scheme\footnote{\url{https://cxc.harvard.edu/proposer/POG/html/chap6.html}} for the Chandra X-ray Telescope as defined in Fig. \ref{grades_def}. All the $3\times3$ pixels of an event are compared to a split threshold $T_{split}$, set to 15 DN (about 10 times of the readout noise). Then the grade values of each of the over-threshold pixels are added together to obtain the total grade of the event. In this way, an event grade from Grade 0 to Grade 255 is assigned to each event.  

In the study of the X-ray performance, three kinds of spectra are mostly used: GAtotal, G0total and G0center. The GAtotal spectrum is extracted from events with all the grades and the total charge in the $3 \times 3$ region of each event is used. If only Grade 0 events are used, we get the G0total spectrum. Furthermore, if we only use the center pixel's charge of each Grade 0 event, we can build the G0center spectrum. Apparently, the G0center spectrum has the best energy resolution, which is affected by the noise of the center pixel only. However, the GAtotal spectrum shows the profile with more counts, making it a better choice under low flux. And the G0total spectrum is sometimes used to keep a balance between the two.  

\subsection{Response to the $^{55}\!\rm{Fe}$ source}
\label{sec_gain}
\begin{figure*}[htbp]
\centering
\includegraphics[width=1.0\textwidth]{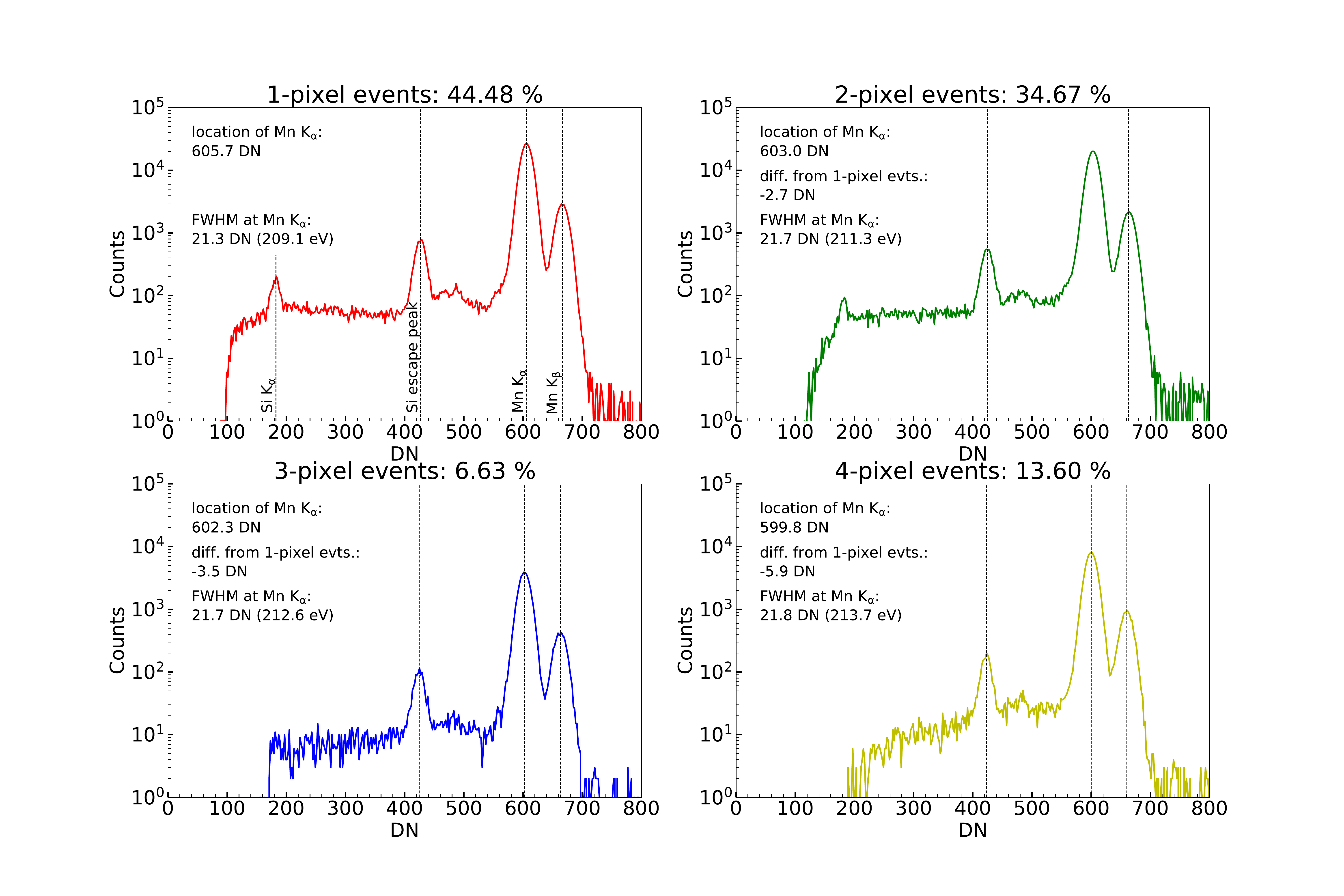}
\caption{The typical spectrum of different types of events of the $^{55}\!\rm{Fe}$ source at $\rm{20 ^{\circ}\!C}$, with the PGA gain register set to 5.0. The total charge in each $3 \times 3$ region is used. Single-pixel events are the Grade 0 events; 2-pixel events contain all the events of Grade 2, 8, 16 and 64; 3-pixel events consist of Grade 10, 18, 72 and 80 events; and 4-pixel events represent all events of Grade 11, 22, 104 and 208. Four lines can be identified: Si $\rm{K_\alpha}$ (1.74 keV), Si escape peak of Mn $\rm{K_\alpha}$ (4.16 keV), Mn $\rm{K_\alpha}$ (5.90 keV), and Mn $\rm{K_\beta}$ (6.49 keV). Single-pixel events give a clearer energy spectrum and a better energy resolution than others.}
\label{spectrum}
\end{figure*}

\begin{figure}[htbp]
\centering
\resizebox{\hsize}{!}{
\includegraphics{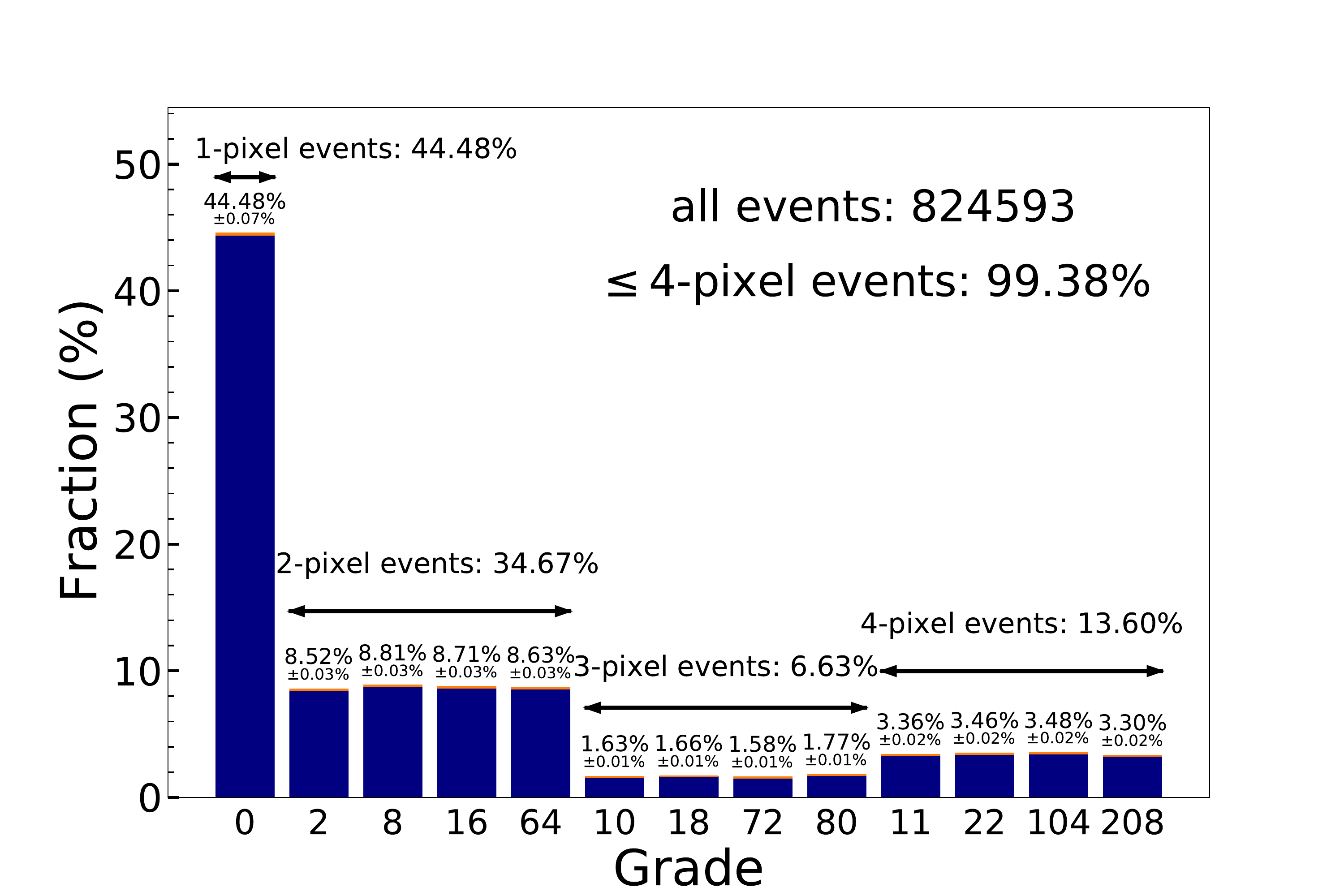}}
\caption{The distribution of grades at $\rm{20 ^{\circ}\!C}$, with the $^{55}\!\rm{Fe}$ source. The PGA gain register is set to 5.0. Single-pixel events take almost half of all events. The relative fraction difference between 2-pixel events is less than 5\%.}
\label{grades_distribution}
\end{figure}

\begin{figure}[htbp]
\centering
\resizebox{\hsize}{!}{
\includegraphics{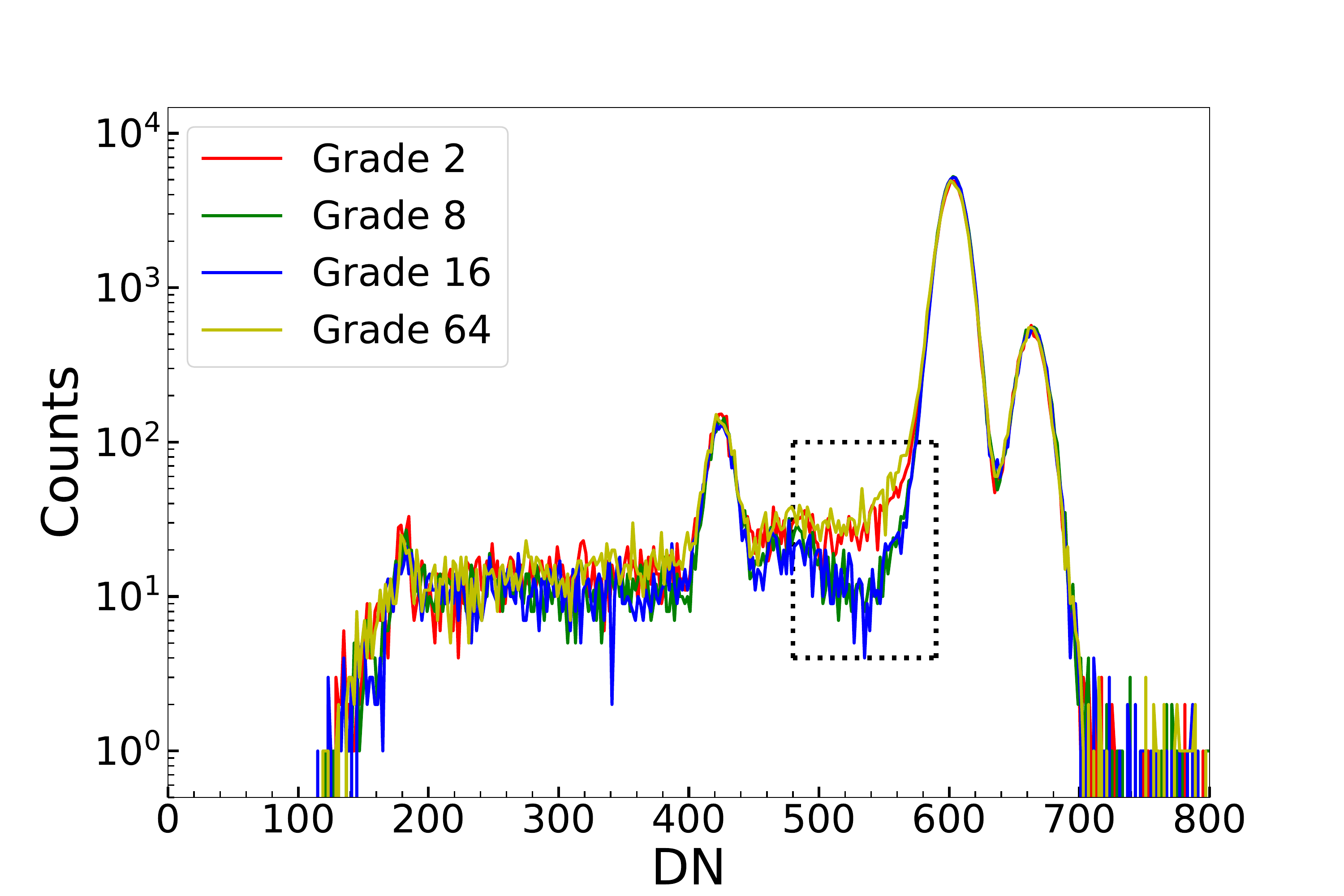}}
\caption{The spectrum of 2-pixel events with different grades at $\rm{20 ^{\circ}\!C}$, with the PGA gain register set to 5.0. There is only a small difference on the spectrum: the spectrum of longitude 2-pixel events (Grade 2/64) has a slightly higher tail (see the spectrum in the black dotted rectangle) than the transverse one (Grade 8/16).}
\label{grades_symmetry}
\end{figure}

An $^{55}\!\rm{Fe}$ X-ray source was used to test the performance of GSENSE\-1516\-BSI. Fig. \ref{spectrum} shows the spectrum of different events of the $^{55}\!\rm{Fe}$ X-ray source at $\rm{20 ^{\circ}\!C}$. The programmable gain amplifier (PGA) will be introduced later and the PGA register is set to 5.0 in this study. The total charge in the $3 \times 3$ region is used. Single-pixel events are the Grade 0 events; 2-pixel events contain all the events of Grade 2, 8, 16 and 64; 3-pixel events consist of Grade 10, 18, 72 and 80 events; and 4-pixel events represent all events of Grade 11, 22, 104 and 208. In the spectrum of single-pixel events, four lines can be identified: Si $\rm{K_\alpha}$ (1.74 keV), Si escape peak of Mn $\rm{K_\alpha}$ (4.16 keV), Mn $\rm{K_\alpha}$ (5.90 keV), and Mn $\rm{K_\beta}$ (6.49 keV). Single-pixel events have a clearer energy spectrum and a better energy resolution than the others. Apparently, the Mn $\rm{K_\alpha}$ peak shift of multi-pixel events indicates the signal loss during electron collection. This charge loss may result from the recombination due to the long diffusion distance and the absorption by the edge of pixels. Fig. \ref{spectrum} shows that 2-pixel events have a peak shift of 2.7 DN ($\sim 7 \rm{e^-}$), around 0.5\% of the peak energy, and 3-pixel and 4-pixel events have slightly larger peak shifts. This result indicates the charge loss of GSENSE\-1516\-BSI is obviously improved over GSENSE\-400\-BSI \citep{narukage2020high}. A more detailed distribution of grades is given in Fig. \ref{grades_distribution}. The asymmetry between longitude 2-pixel events (Grade 2/64) and transverse 2-pixel events (Grade 8/16) has also been reduced significantly compared to GSENSE\-400\-BSI \citep{ling2021correlogram}. The count rates for the longitude events and the transverse one are consistent at a 5\% level. There is only a small difference in the spectra (see Fig. \ref{grades_symmetry}): the spectrum of longitude 2-pixel events (Grade 2/64) has a slightly higher tail than the transverse one (Grade 8/16).            

\begin{figure}[htbp]
\centering
\resizebox{\hsize}{!}{
\includegraphics{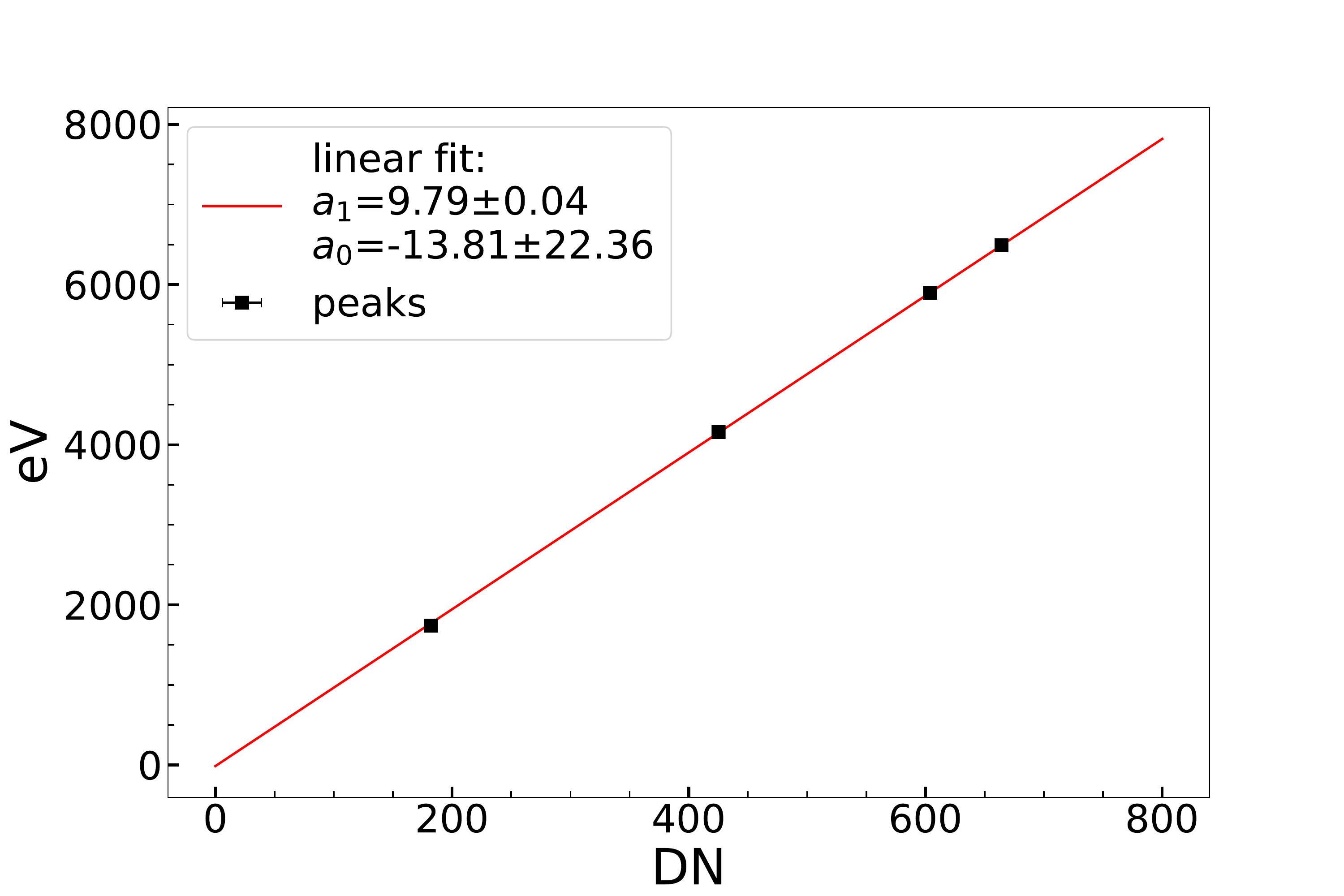}}
\caption{A linear fit is applied to the peaks of the GAtotal spectrum to obtain the conversion gain ($9.79 \pm 0.04$ eV/DN) at $\rm{20 ^{\circ}\!C}$, with the PGA gain register set to 5.0.}
\label{gainfit}
\end{figure}

With the peaks identified in Fig. \ref{spectrum}, we fit the peaks with a linear function
\begin{equation}
y = a_1\times x + a_0, 
\end{equation}
where x is the position of the peaks in DN and y is the energy of the line in eV. Then we can obtain the conversion gain $a_1$. Fig. \ref{gainfit} is the fitting result of the GAtotal spectrum: the conversion gain is 9.79 eV/DN or 2.68 $\rm{e^-}$/DN at $\rm{20 ^{\circ}\!C}$. The full width at half maximum (FWHM) of the Mn $\rm{K_\alpha}$ peak (5.90 keV) is also calculated using a Gaussian fit. Compared to multi-pixel events, single-pixel events (G0total spectrum) achieve the best energy resolution of 209.1 eV (3.5\%) at 5.9 keV. The GAtotal spectrum's energy resolution is deteriorated to 217.2 eV (3.7\%). This resolution is somewhat far from the Fano-limited energy resolution $\sim119$ eV (2.0\%) at 5.9keV \citep{owens2002experimental, fano1947ionization, mazziotta2008electron}. Except for the effect of the readout noise, an additional noise of $\sim 30 \rm{e^-}$ should be introduced to explain the energy resolution. An optical flat field test shows that the PRNU is around 2\%. This value is consistent with the noise needed. This non-uniformity comes from the performance difference between the first stage amplifying capacitors of pixels. The fraction, the conversion gain and the energy resolution of multi-pixel events are summarized in Table \ref{grades}.      

\begin{table}[h!]
\centering
\resizebox{\hsize}{!}{
\begin{tabular}{| l | l | l | l |}
\hline
Event type & Fraction & Gain (eV/DN) &  Resolution (eV)\\ \hline
1-pixel events (G0total) & 44.48\% & $9.77 \pm 0.03$ & $209.1 \pm 1.5$\\ \hline
2-pixel events & 34.67\% & $9.75 \pm 0.02$ & $211.3 \pm 1.1$\\ \hline
3-pixel events & 6.63\% & $9.78 \pm 0.01$ & $212.6 \pm 3.4$\\ \hline
4-pixel events & 13.60\% & $9.82 \pm 0.01$ & $213.7 \pm 1.7$\\ \hline
$>$4-pixel events & 0.62\% & - & - \\ \hline
all events (GAtotal) & 100\% & $9.79 \pm 0.04$ & $217.2 \pm 1.1$\\  
\hline
\end{tabular}}
\caption{The fraction, the gain and the energy resolution (FWHM) of the Mn $\rm{K_\alpha}$ line of different types of events. The temperature is fixed at $\rm{20 ^{\circ}\!C}$. The PGA gain register is set to 5.0. The $T_{split}$ is set to 15 DN.}
\label{grades}
\end{table}

\begin{figure}[htbp]
\centering
\resizebox{\hsize}{!}{
\includegraphics{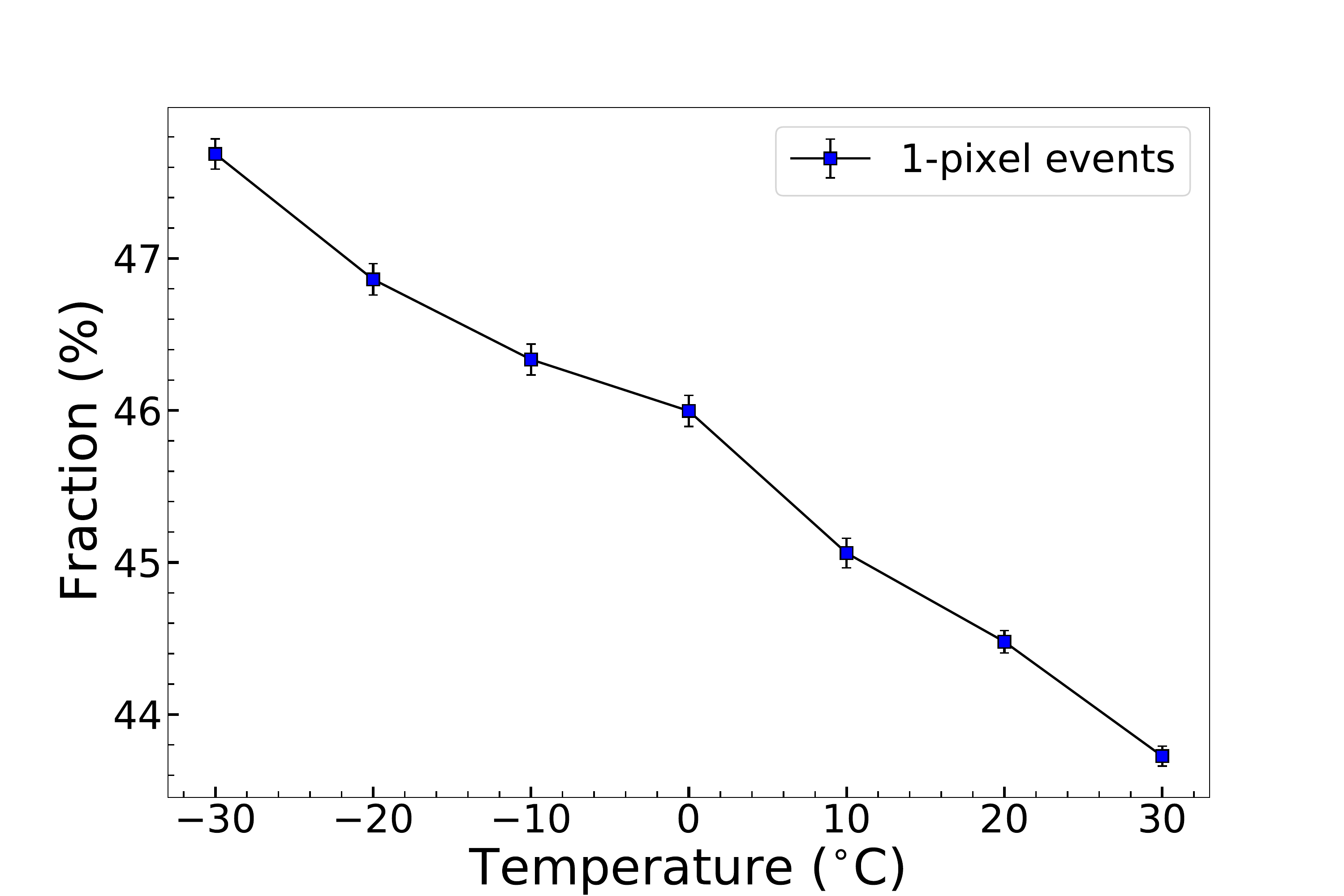}}
\caption{The fraction of 1-pixel events changes with the temperature. This is related to the stronger diffusion of the charge cloud at higher temperature. The $T_{split}$ is kept at 15 DN. The PGA gain register is set to 5.0.}
\label{grades_vs_temp}
\end{figure}

\begin{figure}[htbp]
\centering
\resizebox{\hsize}{!}{
\includegraphics{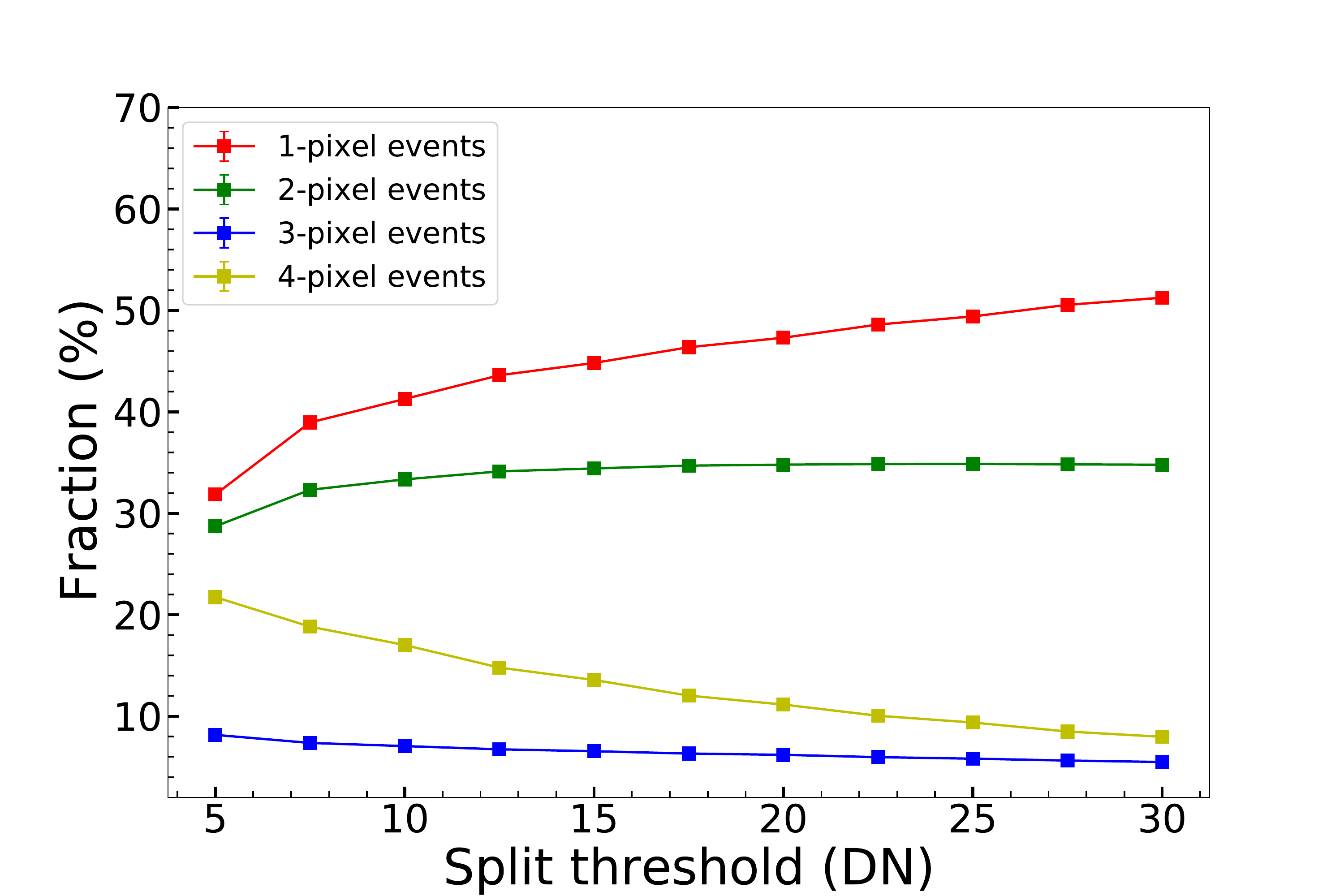}}
\caption{The fraction of different types of events changes with the split threshold at $\rm{20 ^{\circ}\!C}$. The selection of $T_{split}=15$ DN in our data processing is reasonable, because the fraction changes linearly above this value, meaning almost complete exclusion of the noise. The result also indicates a typical single-pixel fraction around 50\%. The PGA gain register is set to 5.0.}
\label{grades_vs_threshold}
\end{figure}

As shown in Table \ref{grades}, single-pixel events make up almost half of all events, much higher than that for GSENSE\-400\-BSI \citep{narukage2020high}. A thicker depletion layer and larger pixel size can account for this improvement. The fraction of grades is also tested at different temperatures and with different split thresholds. As shown in Fig. \ref{grades_vs_temp}, the fraction of 1-pixel events decreases as temperature increasing. This is mainly because of the stronger diffusion of the charge cloud at higher temperature. Fig. \ref{grades_vs_threshold} shows that the fraction of different types of events is affected by the split threshold, especially for 1-pixel and 4-pixel events. We chose $T_{split}$ = 15 DN in our study of grades, because the fraction changes linearly above this value, meaning almost complete exclusion of the noise. Varying $T_{split}$ around this value, the single-pixel fraction changes slightly, but keeps around a typical value of 50\%. However, more accurate measurements are needed to study this fraction more precisely.  

We also calculated the crosstalk of GSENSE\-1516\-BSI, following the method in our previous work \citep{ling2021correlogram}. We used Grade 0 events to draw correlograms, showing the correlation between the DN values of the center pixel and its adjacent pixels. If the DN values of the adjacent pixel are uniformly distributed around 0 DN, then the signal of this pixel comes from the readout noise but not the crosstalk from the center pixel. If the value of an adjacent pixel shows a linear relation with the value of the center pixel, then a linear fit is applied to each correlogram, the slope of which gives the crosstalk. No obvious crosstalk (lower than 0.1\%) is found in this sensor, which is a substantial improvement over GSENSE\-400\-BSI \citep{ling2021correlogram}.

%\subsubsection{Programmable gain amplifiers}
The GSENSE\-1516\-BSI sCMOS has on-chip programmable gain amplifiers (PGAs) to amplify the original signal. The PGA gain register can be set from 0.25 to 7.5 with a step of 0.25, leading to a total of 30 gain levels. The on-chip gain settings would change the value of the conversion gain and the readout noise. The performance of GSENSE\-1516\-BSI is measured under different PGA gain settings at a typical temperature of $\rm{20 ^{\circ}\!C}$, as shown in Table \ref{table_gain_resolution}. The energy resolution at 5.9 keV is given for the GAtotal, G0total and G0center spectra. The G0center spectrum has the best energy resolution because the spectrum is only affected by the noise from one pixel. When the PGA gain register is set to 7.0, we can reach the best energy resolution of 180.1 (3.1\%) at 5.9 keV. Table \ref{table_gain_resolution} also shows that a larger PGA gain value results in a better energy resolution, a lower noise level but a smaller digital full well capacity.    

\begin{table*}[htbp]
\centering
\resizebox{\textwidth}{!}{
\begin{tabular}{| l | l | l | l | l | l | l |}
\hline
Register & Conversion gain & \multicolumn{3}{| c |}{Resolution} & Readout noise & Digital FWC\\ \hline
& GAtotal spectrum & GAtotal spectrum & G0total spectrum & G0center spectrum & & \\ 
& (eV/DN) & (eV) & (eV) & (eV) & ($\rm{e^-}$) & (keV)\\ \hline 
1.0 & $47.41 \pm 0.34 $  & $370.2 \pm 3.1$  & $355.9 \pm 3.0$  & $208.8 \pm 2.0$  & 11.3 & 182.5\\ \hline
2.0 & $24.17 \pm 0.08 $  & $262.1 \pm 2.4$  & $260.2 \pm 3.9$  & $192.3 \pm 1.9$  & 6.7 & 93.1\\ \hline
3.0 & $16.25 \pm 0.07$  & $235.1 \pm 1.4$  & $230.6 \pm 1.6$  & $189.6 \pm 1.6$  & 5.1 & 62.6\\ \hline
4.0 & $12.22 \pm 0.04$  & $220.9 \pm 1.0$  & $214.3 \pm 1.2$  & $188.2 \pm 1.3$  & 4.3 & 47.0\\ \hline
5.0 & $9.79 \pm 0.04$  & $217.2 \pm 1.1$  & $209.1 \pm 1.5$  & $185.7 \pm 1.1$  & 3.9 & 37.7\\ \hline
6.0 & $8.16 \pm 0.02$  & $213.1 \pm 0.8$  & $203.9 \pm 0.8$  & $184.9 \pm 1.0$  & 3.5 & 31.4\\ \hline
7.0 & $7.01 \pm 0.02$  & $212.3 \pm 1.1$  & $202.4 \pm 1.4$  & $180.1 \pm 1.1$  & 3.3 & 27.0\\ 
\hline
\end{tabular}}
\caption{The conversion gain, the energy resolution (FWHM) of Mn $\rm{K_\alpha}$, the readout noise and the digital full well capacity changes with the register value. A larger PGA gain value results in a better energy resolution, a lower noise level but a smaller digital full well capacity. The energy resolution at 5.9 keV is given for the GAtotal, G0total and G0center spectra. The G0center spectrum has the best energy resolution because the spectrum is only affected by the noise from one pixel. The temperature is fixed at a typical value of $\rm{20 ^{\circ}\!C}$.}
\label{table_gain_resolution}
\end{table*}

\begin{figure}[htbp]
\centering
\resizebox{\hsize}{!}{
\includegraphics{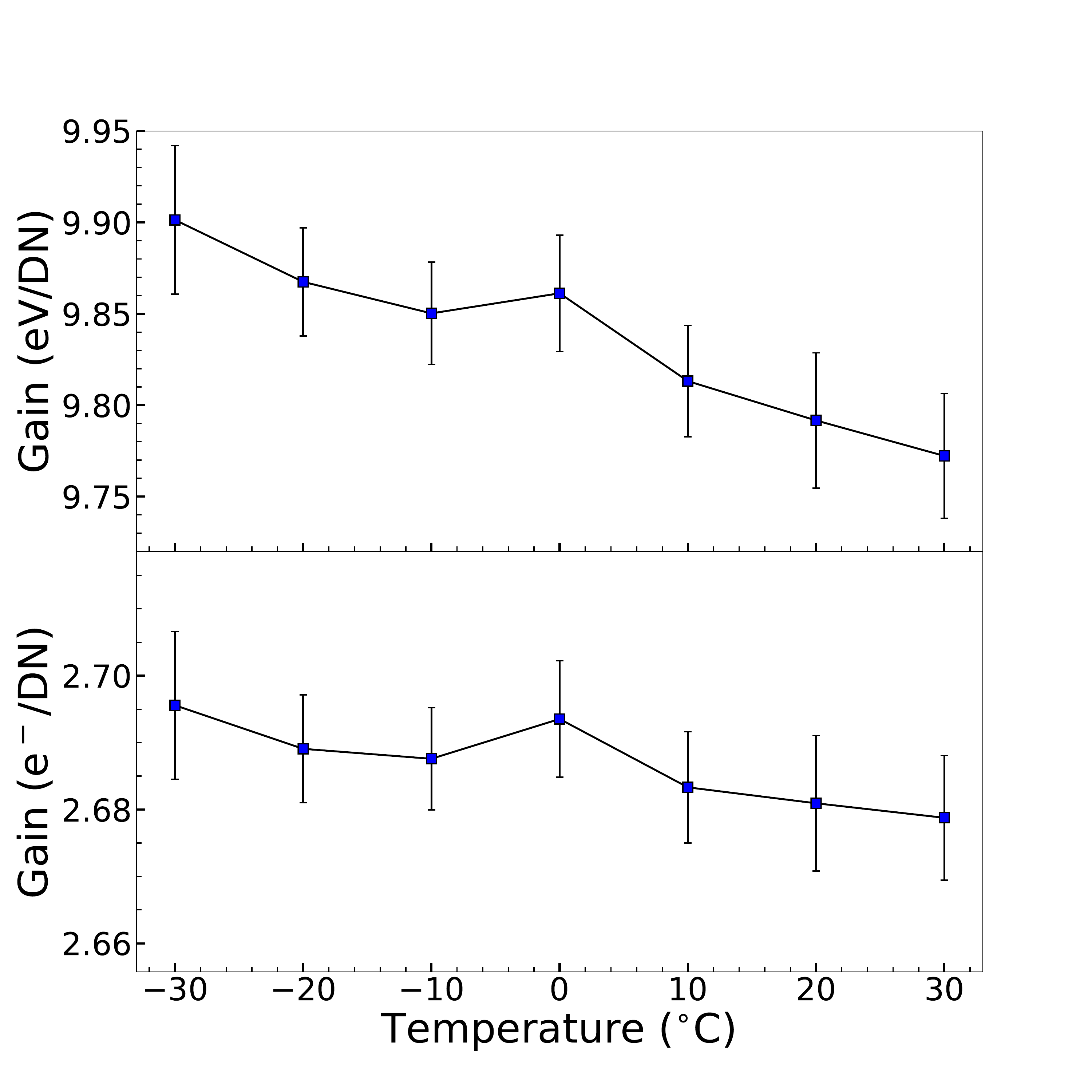}}
\caption{The conversion gain (eV/DN at top and $\rm{e^-}/DN$ at bottom) at different temperatures. The conversion gain of GAtotal spectrum is slightly affected by temperature (top). This is mainly ascribed to the temperature-dependent pair creation energy \citep{mazziotta2008electron}. Apart from this factor, the conversion gain from DN to $\rm{e^-}$ (bottom) shows less dependency on temperature.The PGA gain register is set to 5.0.}
\label{gain_all_vs_temp}
\end{figure}

%\subsubsection{Influence of temperature}
Temperature also influences the X-ray performance of GSENSE\-1516\-BSI. As shown in Fig. \ref{grades_vs_temp}, the fraction of 1-pixel events decreases as temperature increasing. Apart from this, Fig. \ref{gain_all_vs_temp} (top) shows that the conversion gain of GAtotal spectrum is also slightly affected by temperature. This is mainly ascribed to the temperature-dependent pair creation energy \citep{mazziotta2008electron}. Apart from this factor, the conversion gain from DN to $\rm{e^-}$ (see the bottom pattern of Fig. \ref{gain_all_vs_temp}) shows less dependency on temperature. The energy resolution also varies with temperature, mostly following the change of the readout noise (see Table \ref{table_gain_resolution_2}). From $\rm{-30 ^{\circ}\!C}$ to $\rm{30 ^{\circ}\!C}$, GSENSE\-1516\-BSI retains a good energy resolution, enabling its potential applications in various environments. Although some hot pixels may still appear at high temperatures, the small quantity making their influence on the spectrum negligible. Therefore, the energy resolution is slightly better at room temperature, mostly due to the lower noise. The performances at different temperatures are summarized in Table \ref{table_gain_resolution_2}.

\begin{table*}[htbp]
\centering
\resizebox{\textwidth}{!}{
\begin{tabular}{| l | l | l | l | l | l | l |}
\hline
Temperature & Conversion gain & \multicolumn{3}{| c |}{Resolution} & Readout noise & Dark current\\ \hline
& GAtotal events & GAtotal events & G0total events & G0center events & & \\ 
($\rm{^{\circ}\!C}$) & (eV/DN) & (eV) & (eV) & (eV) & ($\rm{e^-}$) & ($\rm{e^-}$/pixel/s)\\ \hline 
-30 & $9.90 \pm 0.04 $  & $235.7 \pm 1.5$  & $219.6 \pm 1.4$  & $192.2 \pm 2.0$  & 3.6 & 0.018\\ \hline
-20 & $9.87 \pm 0.03 $  & $231.9 \pm 1.3$  & $220.8 \pm 1.5$  & $191.1 \pm 1.9$  & 3.6 & 0.032\\ \hline
-10 & $9.85 \pm 0.03$  & $228.6 \pm 1.3$  & $216.1 \pm 1.6$  & $189.9 \pm 1.6$  & 3.5 & 0.091\\ \hline
0 & $9.86 \pm 0.03$  & $224.8 \pm 1.2$  & $214.7 \pm 1.5$  & $187.4 \pm 1.3$  & 3.5 & 0.31\\ \hline
10 & $9.81 \pm 0.03$  & $219.4 \pm 1.0$  & $209.7 \pm 1.2$  & $187.2 \pm 1.1$  & 3.4 & 1.6\\ \hline
20 & $9.79 \pm 0.04$  & $217.2 \pm 1.1$  & $209.1 \pm 1.5$  & $185.7 \pm 1.0$  & 3.3 & 7.4\\ \hline
30 & $9.77 \pm 0.03$  & $215.3 \pm 1.0$  & $209.2 \pm 1.0$  & $185.2 \pm 1.1$  & 3.2 & 36\\
\hline
\end{tabular}}
\caption{The conversion gain, the energy resolution (FWHM) of Mn $\rm{K_\alpha}$, the readout noise and the dark current changes with the temperature. The energy resolution at 5.9 keV is given for the GAtotal, G0total and G0center spectra. The resolution varies with the temperature, mostly following the change of the readout noise. The PGA gain register is set to a typical value of 5.0.}
\label{table_gain_resolution_2}
\end{table*}

\subsection{Response to X-ray continuum spectrum}
\begin{figure}[htbp]
\centering
\resizebox{\hsize}{!}{
\includegraphics{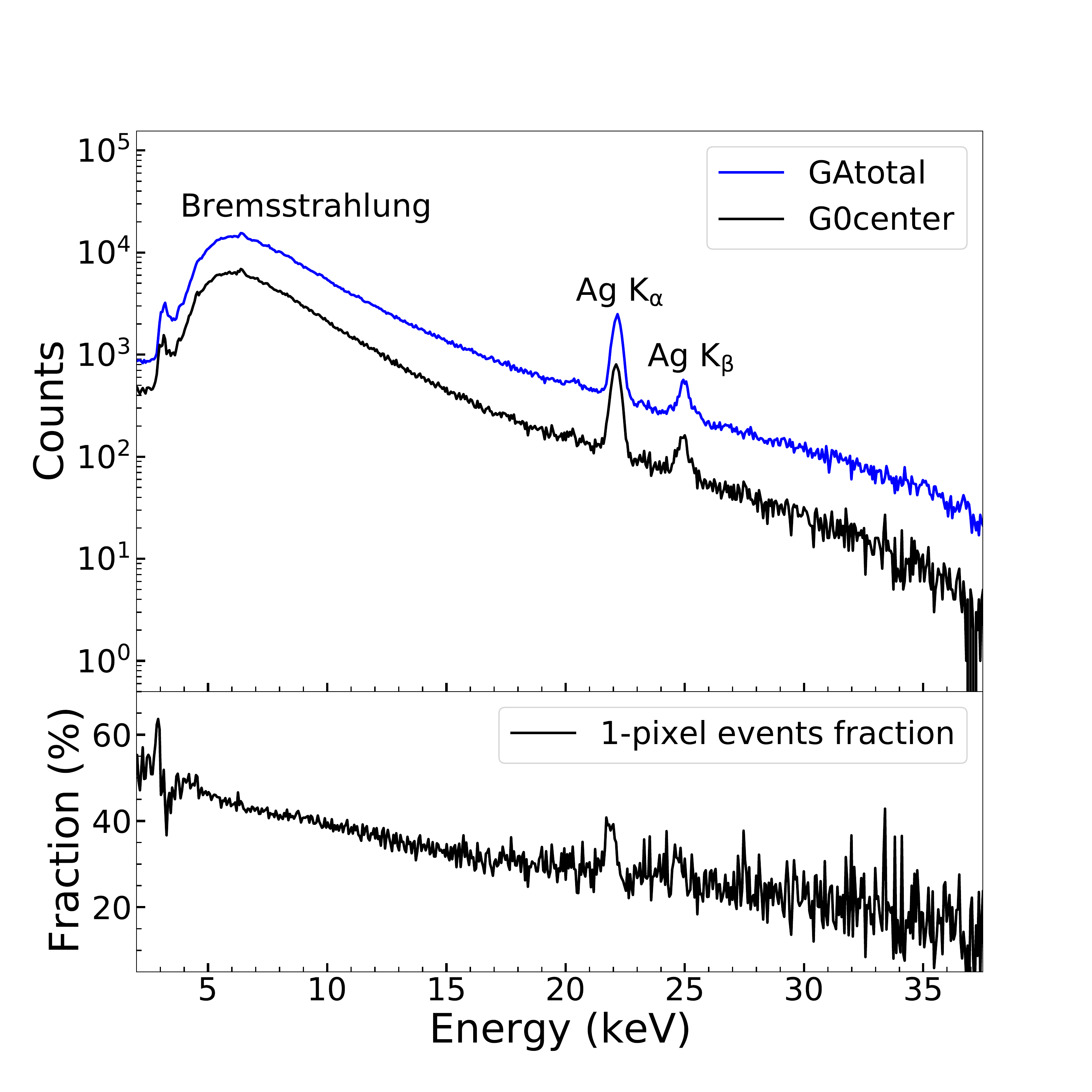}}
\caption{Top: The GAtotal and G0center spectra from the X-ray tube at $\rm{20 ^{\circ}\!C}$. The continuum from Bremsstrahlung radiation and the two peaks of Ag $\rm{K_\alpha}$ and Ag $\rm{K_\beta}$ can be seen clearly. Bottom: The fraction of 1-pixel events decreases with increasing energy, which is due to the diffusion of larger primary electron clouds produced by higher energy photons. The $T_{event}$ is set to 50 DN and the $T_{split}$ is kept at 15 DN. The PGA gain register is set to 5.0.}
\label{x_ray_tube_spec}
\end{figure}

We used an AMPTEK mini X-ray tube to test the broadband spectrum response of GSENSE\-1516\-BSI in this work. The quantum efficiency can be estimated from the epitaxial thickness of 10 $\rm{\mu m}$. The high voltage of the tube is set to 40keV, which is slightly higher than the digital full well capacity at a PGA gain of 5.0. The spectrum in Fig. \ref{x_ray_tube_spec} proves that this sCMOS is capable of detecting X-ray photons up to 37 keV. The continuum from Bremsstrahlung radiation and the two peaks of Ag $\rm{K_\alpha}$ and Ag $\rm{K_\beta}$ can be seen clearly. The fraction of 1-pixel events decreases with increasing energy, which is due to the diffusion of larger primary electron clouds produced by higher energy photons. The $T_{event}$ is set to 50 DN and the $T_{split}$ is kept at 15 DN in this test.

%The BSI sCMOS is also tested under the illumination of a Ag targert X-ray tube source. The spectrum in Fig. \ref{x_ray_tube_spec} proves that this sCMOS is capable of detecting X-ray photons up to 35 keV. The continuum from Bremsstrahlung radiation and the two peaks of Ag ($\rm{K_\alpha}$ and $\rm{K_\beta}$) can be seen clearly. The rate of 1-pixel events decreases with increasing photon energy because of the larger charge cloud produced by an incoming high energy photon. 

\subsection{Detecting low energy X-rays}
\begin{figure}[htbp]
\centering
\resizebox{\hsize}{!}{
\includegraphics{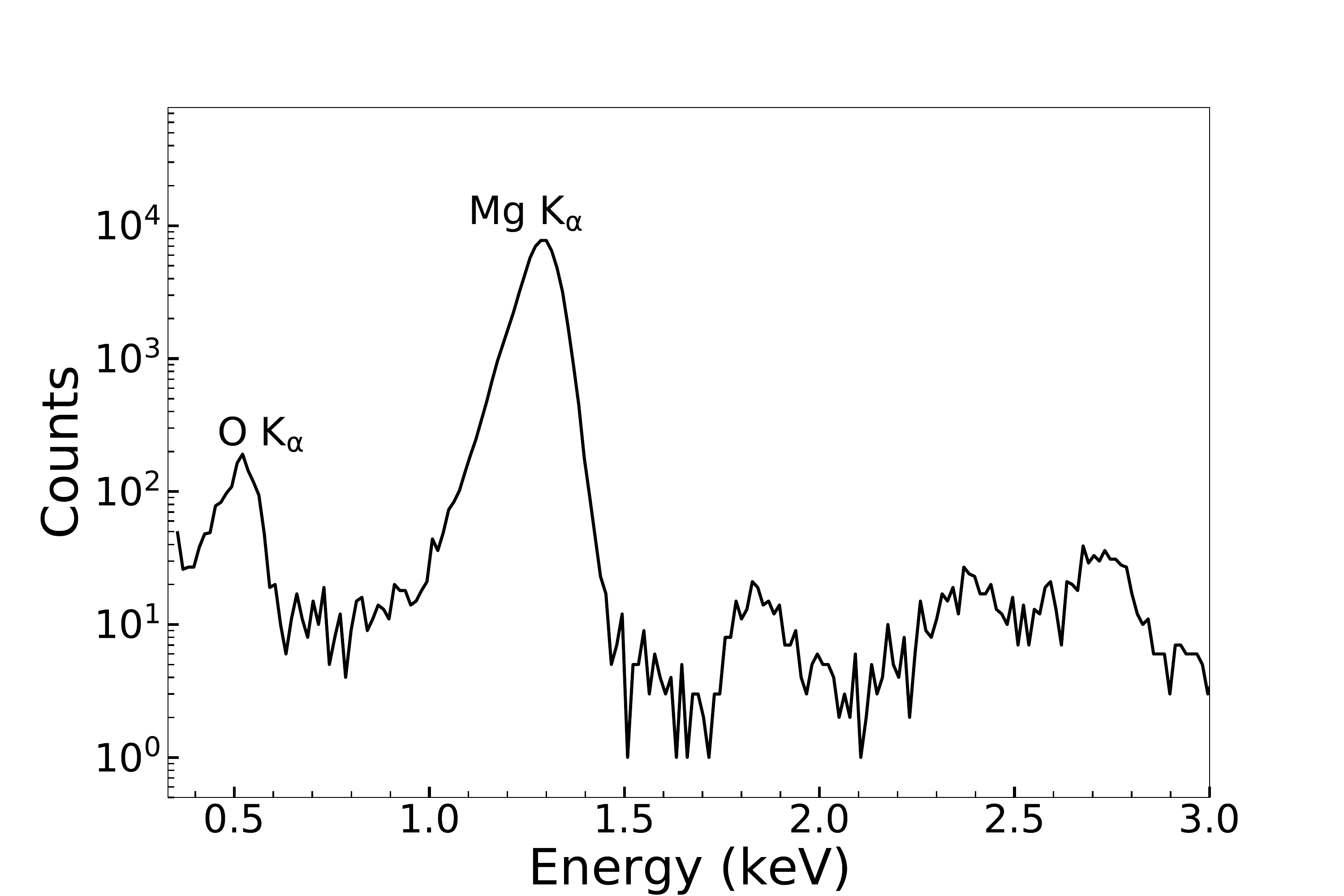}}
\caption{Spectrum from the Mg target at $\rm{-30 ^{\circ}\!C}$. The 525 eV O $\rm{K_{\alpha}}$ and 1254 eV Mg $\rm{K_{\alpha}}$ lines can be identified clearly. The G0center spectrum is plotted to show the emission line clearly. The $T_{event}$ is set to 40 DN and the $T_{split}$ is kept at 15 DN. The PGA gain register is set to 7.5.}
\label{low_energy_spec}
\end{figure}

The lower energy threshold of the sensor is limited by the entrance window and the readout noise. In this research, we employed an X-ray-illuminated Magnesium target to produce the emission line. The spectrum is shown in Fig. \ref{low_energy_spec}, where the $T_{event}$ is set to 40 DN and the $T_{split}$ is kept at 15 DN. The G0center spectrum is plotted to show the emission line clearly. The 525 eV O $\rm{K_{\alpha}}$ and 1254 eV Mg $\rm{K_{\alpha}}$ lines can be identified clearly, showing that the sensor is capable of detecting low energy X-ray photons down to $\sim$500 eV.

\section{Conclusions}
\label{conclusions}
We proposed a dedicated sCMOS detector aiming at astronomical utilization in collaboration with Gpixel Inc. The product, namely GSENSE\-1516\-BSI, was successfully produced in 2019. In this work, we present the test results of the performance of this X-ray sensor. The dark current of this sensor is 7.4 $\rm{e^-}$/pixel/s at $\rm{20 ^{\circ}\!C}$, and 0.018 $\rm{e^-}$/pixel/s at $\rm{-30 ^{\circ}\!C}$. The FPN and the readout noise are 3.9 $\rm{e^-}$ and 3.3 $\rm{e^-}$ at $\rm{20 ^{\circ}\!C}$, respectively. Both types of noise show a small increase at a low temperature. The energy resolution reaches 180.1 eV (3.1\%) at 5.90 keV for single-pixel events (G0center) and 212.3 eV (3.6\%) for all split events (GAtotal). The fraction of single-pixel events is around 50\% and is slightly affected by temperature and split threshold. The continuum X-ray spectrum measurement shows that this sensor is able to response to X-rays from 0.5 keV to 37 keV. Our test results show that the GSENSE\-1516\-BSI sensor has excellent performance in the detection of soft to hard X-ray photons, with some features comparable to or even better than X-ray CCDs commonly used. It therefore has a great potential for astronomical applications, especially for soft X-ray detection. 

Scientific CMOS sensors can be further upgraded and adapted for various application requirements. For the detection of low energy X-ray photons, the thickness of the entrance window of the sensor can be reduced to several nanometers \citep{harada2020high}. To improve the high energy quantum efficiency, we are proposing a sCMOS sensor with a thick depletion layer of several hundred micrometers. Developing sCMOS sensors with larger format, higher readout speed, lower readout noise and thicker depletion layer is also under consideration. Such sensors will be useful for the next generation of large-area X-ray missions, e.g., the Lynx mission. Compared with CCD, sCMOS sensors can operate at a much higher readout speed: the maximum frame rate of GSENSE\-1516\-BSI is around 100 Hz. This fast readout feature makes the sCMOS sensors more suitable for applications in time-domain astronomy (e.g., fast photometry) and wide-field sky surveys. Another advantage of sCMOS sensors is that there is no charge transfer process during the readout. For optical imaging of bright sources, the images are free from the influence of readout trails and charge residues. For X-ray imaging and spectroscopic application, sCMOS sensors can provide generally high-quality data comparable to those acquired with CCD sensors. In addition, there is no Out Of Time (OOT) events. sCMOS sensors have a similar noise level and energy resolution to CCD sensors at a high working temperature. This advantage helps lower the power consumption and simplify the design of the cooling system, making it easier and more cost effective to build large area focal plane detectors. In conclusion, thanks to its advantages of low cost, fast readout speed, high working temperature and high radiation tolerance, the research and development of sCMOS sensors are booming, and the future of their applications is bright.

\begin{acknowledgments}
This work is supported by the National Natural Science Foundation of China (grant no. 12173055) and the Chinese Academy of Sciences (grant no. XDA15310100, XDA15310300, XDA15052100).
\end{acknowledgments}






\bibliography{references.bib}{}
\bibliographystyle{aasjournal}



\end{document}